\documentclass[journal]{IEEEtran}

\usepackage[english]{babel}
\usepackage{epstopdf}
\usepackage{amsmath}
\usepackage{multirow}
\usepackage{epsfig}
\usepackage{mathrsfs}
\usepackage{amssymb}
\usepackage{comment}
\usepackage{array}
\usepackage{blkarray}
\usepackage{enumerate}
\usepackage{url}
\usepackage{chemarrow}
\usepackage{makecell}

\makeatletter
\newif\if@restonecol
\makeatother

\usepackage[lined,ruled,linesnumbered]{algorithm2e}

\newtheorem{theorem}{Theorem}
\newtheorem{definition}[theorem]{Definition}

\usepackage[dvipsnames]{xcolor}
\definecolor{mycolour}{rgb}{0,0,0}

\usepackage[tight,footnotesize]{subfigure}

\begin{document}

\title{Neural Network Tomography}




%

\author{Liang~Ma,~\IEEEmembership{Member,~IEEE,}
	Ziyao~Zhang,~\IEEEmembership{Student Member,~IEEE,} and~Mudhakar~Srivatsa,~\IEEEmembership{Senior Member,~IEEE}}


\maketitle
\begin{abstract}
	Network tomography, a classic research problem in the realm of network monitoring, refers to the methodology of inferring unmeasured network attributes using selected end-to-end path measurements. In the research community, network tomography is generally investigated under the assumptions of known network topology, correlated path measurements, bounded number of faulty nodes/links, or even special network protocol support. The applicability of network tomography is considerably constrained by these strong assumptions, which therefore frequently position it in the theoretical world. In this regard, we revisit network tomography from the practical perspective by establishing a generic framework that does not rely on any of these assumptions or the types of performance metrics. Given \emph{only} the end-to-end path performance metrics of sampled node pairs, the proposed framework, \emph{NeuTomography}, utilizes deep neural network and data augmentation to predict the unmeasured performance metrics via learning non-linear relationships between node pairs and underlying unknown topological/routing properties. In addition, NeuTomography can be employed to reconstruct the original network topology, which is critical to most network planning tasks. Extensive experiments using real network data show that comparing to baseline solutions, NeuTomography can predict network characteristics and reconstruct network topologies with significantly higher accuracy and robustness using only limited measurement data.
\end{abstract}

\IEEEpeerreviewmaketitle

\section{Introduction}
\label{intro}

Accurate and timely knowledge of network states (e.g., delays and congestion levels on individual paths) is essential for various network operations such as route selection, resource allocation, fault diagnosis, and service migrations. Directly measuring the path quality with respect to (w.r.t.) each individual node pair, however, is costly and not always feasible due to the large traffic overhead in the measurement process and the lack of protocol support at internal network elements for making such measurements \cite{LoPresti02}.
Hence, such limitations motivate the need for \emph{indirect} approaches, where we infer the network states of interest by measuring the performance along selected paths w.r.t. a subset of node pairs.

Depending on the granularity of observations, indirect approaches can be classified as \emph{hop-by-hop} or \emph{end-to-end} approaches. The former rely on special diagnostic tools such as \emph{traceroute}, \emph{pathchar} \cite{Downey99}, \emph{Network Characterization Service (NCS)} \cite{Jin01HPDC}, or \emph{Time Series Latency Probes (TSLP)} \cite{Dhamdhere18SIGCOMM} to reveal fine-grained performance metrics of individual links by sending active probes. Specifically, Traceroute reports delay for each hop on the probed path by gradually increasing the time-to-live (TTL) \cite{Lai00SIGCOMM} field of probing packets. Its refinement, pathchar, returns hop-by-hop delays and loss rates. Later advancements, NCS and TSLP, return capacities on each link. While providing fine-grained information, the above tools require that Internet Control Message Protocol (ICMP) be supported at each internal node. Even then, they cause extra load and suffer from inaccuracies caused by different priorities of ICMP/data packets. 
In risk-sensitive applications, security policies may even block such measurements.

Alternatively, the end-to-end approach provides a solution that does not require the cooperation of internal network elements or the equal treatment of control/data packets. It relies on end-to-end path performance metrics (e.g., path delays or bandwidths) experienced by data packets to infer the unknown network information using \emph{network tomography}. Formally, network tomography \cite{Coates02} refers to the methodology of inferring unmeasured network characteristics via measuring the end-to-end performance of selected paths.

As a generic case, network tomography can be used to infer path performance metrics for any unmeasured node pair, which covers the case of link performance metric inference as links essentially correspond to $1$-hop paths. 
In real networks, the path performance metrics of interest can be \emph{additive}, i.e., the combined metric over multiple links is the sum of individual link metrics, or \emph{non-additive}, i.e., the path metric is a non-linear function of the involved link metrics.
For instance, delays are additive, while a multiplicative metric (e.g., packet delivery ratio) can be expressed in an additive form using function $\log (\cdot)$. By contrast, path congestion levels are determined by the worst performed link on this path, thus non-additive. The general goal of network tomography is to utilize the path measurements between a small subset of node pairs to infer the path performance metrics of the remaining unmeasured node pairs.

Existing work on network tomography emphasizes extracting as much network information as possible from available measurements. However, past experience shows that network tomography mostly exists in the theoretical level under strong assumptions, thus difficult to be applied to real network monitoring tasks. For example, prior works \cite{GurewitzSidi01,Chen04,Chen07,Lawrence06,Mahajan02,Ma13IMC,Ma14ToN,Ma13ICDCS,Liu15MILCOM,He15Sigmetrics,Tati14ICDCS,Gao14ICNP,Gao16ToN,Gao14ICDCS,Li18ToN,Ren16INFOCOM,Dong17ToN,Ma13GLOBECOM,Ma14INFOCOM,He16Infocom,He17ToN,Ma14IMC,Ma17ToN,Bartolini17INFOCOM,Ma15Performance,He16ICDCS,He17DistributedLA,Tootaghaj18Performance,Galesi18ICDCS,Ghita10INFOCOM,Zhang14SIGCOMM,Ghita10IMC} almost always require the precise knowledge of the network topology and \cite{Bu02,Duffield00,Duffield03,Duffield06TInfo,Liu15MILCOM} even assume that the network topology follows a tree structure. Unfortunately, in real applications, network topologies are frequently concealed from others for security reasons. Furthermore, special routing mechanism such as source routing is needed in \cite{Ma13IMC,Ma14ToN,Ma13ICDCS,Gao16ToN,Gao14ICNP,Gao14ICDCS,Li18ToN,Ren16INFOCOM,Dong17ToN,Ma13GLOBECOM,Ma14INFOCOM,He16Infocom,He17ToN} and multicast is employed in \cite{Lawrence06,Bu02,Duffield00,Duffield03,Duffield06TInfo}; however, practical networks usually block the support of such strong routing requirement. Moreover, regarding additive performance metrics, besides the end-to-end path performance metric, the nodes/links involved in this path are often required for the performance inferring task. While for non-additive performance metrics, the maximum number of problematic nodes/links is usually imposed as a common constraint \cite{Ma14IMC,Ma17ToN,Bartolini17INFOCOM,Ma15Performance,He16ICDCS,He17DistributedLA,Tootaghaj18Performance,Galesi18ICDCS}. Finally, little is known for a  solution that simultaneously addresses both additive and non-additive tomography problems.

In this paper, we establish a generic and lightweight tomography framework that \emph{removes all} above assumptions and constraints, thus applicable to most practical network setting. The input to our tomography framework is only a set of end-to-end path measurements w.r.t. some node pairs, and the output is the predicted path performance metrics for all unmeasured node pairs. For each input data point, the \emph{only} available information is the starting/terminating nodes and their corresponding path performance metric. 
The proposed framework, called \emph{NeuTomography}, is based on deep neural networks \cite{Goodfellow16book}, which learn the non-linear relationship between node pairs and their path performance metrics. 

Comparing to existing tomography solutions, NeuTomography is generic and easily applicable in that it does not require additional network knowledge or rely on specific performance metric type (additive or non-additive). Moreover, since the given measured node pairs potentially can be any subset of all node pairs in the network, i.e., there may exist measurement bias, we further propose Path Augmented Tomography (PAT) that proactively constructs additional input data by estimating the performance bounds of unmeasured node pairs using the given path measurements. Extensive experiments via both Rocketfuel \cite{RocketFuel} and CAIDA's ITDK \cite{CAIDA} network data show that by measuring only $30\%$ node pairs, NeuTomography is able to accurately predict the path performance of the rest $70\%$ node pairs, with the mean absolute percentage error (MAPE) as small as $2\%$, and PAT further reduces MAPE by up to $50\%$; such results are orders of magnitude improvement over benchmarks. Finally, although we are not given any information about the network topology, NeuTomography provides a solution to efficiently reconstruct the network topology with different granularities utilizing only the given end-to-end node pair measurements, 
thus revealing more insights to network operators for resource optimizations.

\subsection{Further Discussions on Related Work}
\label{subsec:relatedWork}

Network tomography can be categorized into passive \cite{Medina02SIGCOMM} and proactive tomography \cite{Chen04}. Passive tomography refers to a technology of inferring network performance metrics by passively observing the existing traffic attributes \cite{Zhang09SIGCOMM}. 
However, passive tomography requires additional assumptions, e.g.,  correlated performance metrics, to assist the inference task, thus limiting its applicability. In contrast, 
active tomography proactively measures some performance metrics; it is more useful in practical network monitoring tasks, and therefore is the focus of this paper. 


For active tomography, the most important branch is identifying additive performance metrics. Under the assumption of \emph{known} network topology and node/link involvement in each measurement path, the problem is formulated as solving a system of linear equations. Yet, even under such assumptions, it is frequently impossible to uniquely identify all unmeasured link metrics from path measurements
because the linear system is not always invertible \cite{GurewitzSidi01,Chen04,Chen07}, and statistical approaches \cite{Duffield00,LoPresti02,XiaTse06} and multicast \cite{Adam_Towsley00,Castro04,Lawrence06,Bu02} are needed to estimate the link metric distributions.
When all but $k$ link metrics are \emph{zero}, compressive sensing techniques are used to identify the $k$ non-zero link metrics \cite{Firooz10,Xu11,Zhang09SIGCOMM}. With additional assumptions of controllable routing, 
\cite{Gopalan11} derives necessary and sufficient conditions on the network topology for identifying all link metrics, given that monitors can measure any cycles. A similar study in \cite{Alon11} quantifies the minimum number of measurements needed to identify a broader set of link metrics. 
Moreover, \cite{Bejerano03INFOCOM,Kumar06JSAC,Horton03IMC} develop measuring vantage placement algorithms for performing efficient path measurements. Since routing along cycles is typically prohibited, these methods are not widely applicable. If only measuring cycle-free paths are allowed, then \cite{Ma13IMC,Ma14ToN} establish the necessary and sufficient topological conditions for link metric determination. Yet, such cycle-free controllable source routing still has limited support in practice.


When the performance metrics are non-additive, additional constraints are typically imposed.
Under the assumption that multiple simultaneous failures happen with negligible probability, \cite{Lai00SIGCOMM,Carter96Performance,Jain03ToN,Alouf01INFOCOM,Bejerano03INFOCOM,Horton03IMC} target to detect and localize the bottleneck in the network. 
To improve the resolution in characterizing failures, range tomography in \cite{Zarifzadeh12IMC} not only localizes the failure, but also estimates its severity (e.g., congestion level). These works, however, ignore the fact that multiple failures occur more frequently than one may imagine \cite{Markopoulou04infocom}. To address the issue of multiple failures, 
\cite{Duffield03,Duffield06TInfo,Kompella07infocom,Zeng12CoNEXT,Dhamdhere07CoNEXT,Huang08ccr} attempt to find the minimum set of network elements whose failures explain the observed path states. 
In a Bayesian formulation, \cite{Nguyen07infocom,Ghita11CoNEXT} estimate the failure probabilities of different links. 
For the case of binary performance metrics (failed or normal),
if the number of failed links is upper bounded by $k$ and the measuring vantages can probe arbitrary cycles or paths, \cite{Ahuja08,Cho14Elsevier,Ma15Performance} focus on placing measuring vantages and constructing measurement paths to localize a given number of failures. Furthermore, in \cite{Ma14IMC,Ma17ToN,Bartolini17INFOCOM,He16ICDCS,He17DistributedLA,Tootaghaj18Performance,Galesi18ICDCS}, efficient testing conditions and algorithms are proposed to quantify the capability of localizing node failures.
However, for arbitrary valued non-additive link metrics, few positive results are known. In this regard, we build the neural-network-based tomography framework for such general non-additive performance metrics.

\subsection{Summary of Contributions}

Our contributions are four-fold:

\emph{1)} We propose for the \emph{first} time a deep neural-network-based generic and lightweight tomography framework (\emph{NeuTomography}) for network monitoring tasks using only end-to-end path measurements of a subset of node pairs without requiring any additional assumptions on the network.

\emph{2)} We build algorithm Path Augmented Tomography (PAT) to improve the performance prediction accuracy using estimated performance bounds as the augmented input data.

\emph{3)} Although no prior knowledge about the network topology is given, we establish one method using the proposed tomography framework to reconstruct the network topology. 

\emph{4)} Extensive experiments using real data confirm the high accuracy of NeuTomography in predicting path performance metrics of unmeasured node pairs; 
the reconstructed network topologies also exhibit small errors. Such results are orders of magnitude improvement over baseline solutions.

\vspace{.5em}
The rest of the paper is organized as follows. Section~\ref{Sect:ProblemFormulation} formulates the problem. Section~\ref{sec:problemChallenges} discusses the challenges. Section~\ref{sec:NNmodel} presents the proposed tomography framework. Real network data are employed in Section~\ref{sec:experiments} for evaluations. Finally, Section~\ref{sec:Conclusion} concludes the paper.


\section{Problem Statement}
\label{Sect:ProblemFormulation}
In this paper, we consider the most challenging and practical problem setting for network tomography, where strong and/or unrealistic assumptions in prior works are removed. 

\subsection{Path Performance Metric}
\label{subsec:pathPerformanceMetric}
For the end-to-end path performance metric w.r.t. a node pair, it can be broadly classified into two categories:

\noindent \emph{1) Additive performance metrics:} Path performance metric is additive if the combined metric over a path is the sum of all involved individual link metrics. For instance, 
  \begin{itemize}
  \item path length (e.g., number of hops on the path) and path delay are directly additive;
  \item some statistical attributes are additive, e.g., path jitter is the combination of individual link jitters if all link latencies are mutually independent;
  \item  multiplicative path metric (e.g., packet delivery ratio) can be expressed in an additive form of individual link metrics by using the $\log(\cdot)$ function.
  \end{itemize}
\noindent \emph{2) Non-additive performance metrics:} Path performance metric is non-additive if the path metric is not the sum of all involved individual link metrics. For instance,
    \begin{itemize}
    \item binary path status (normal or failed), i.e., the end-to-end path is normal if all involved links are normal, and failed if there exists one failed link on this path;
    \item path congestion level and bandwidth, i.e., they are determined by the most problematic link on this path.
    \end{itemize} 

In this paper, we do \emph{not} impose any constraints on the type of path performance metrics. Our goal is to establish a generic and practical network tomography framework that is capable of handling \emph{any} performance metrics of interest.

\subsection{Problem Input: Path Measurements}
\label{subsec:problemInput}

For a network $\mathcal{G}$ with $n$ nodes, there are $n \choose 2$ node pairs. Let $V$ denote the set of nodes in $\mathcal{G}$ ($|V|=n$) and set $\{v_i,v_j\}$ ($v_i,v_j\in V$, $v_i\neq v_j$) a node pair. Then
$T=\big\{\{v_i,v_j\}\big\}_{v_i,v_j\in V, v_i\neq v_j}$ 
is a set containing all node pairs in $V$ (i.e., $|T|={n \choose 2}$). 
For a given performance metric of interest (any metric as discussed in Section~\ref{subsec:pathPerformanceMetric}), suppose we are given the measured end-to-end path performance metric associated with each node pair in a subset $S$ of $T$, i.e., $S\subset T$. 
Then we explore how to infer the unmeasured path information as accurately as possible \emph{purely} based on this available set of path measurements. 
To make our problem more practical and applicable to real networks, we do not constrain the way to obtain the set of path measurements, or make additional assumptions. Specifically,

 1) For any node pair $\{v_i,v_j\}\in S$, we only know the performance metric of the end-to-end path between $v_i$ and $v_j$. Besides the end-points $v_i$ and $v_j$, we have \emph{no} knowledge on which other nodes are on this path. 

 2) The path between node pair $\{v_i,v_j\}$ in $S$ is constructed via the underlying routing protocol(s), which are \emph{unknown}. 

 3) We do \emph{not} know the network topology or even the number of links in the network. 

 4) \emph{No} constraints are imposed on how the given node pair set $S$ is obtained; therefore, potentially $S$ can be any subset of $T$. Hence, the path metric distribution associated with node pairs in $S$ might be different from the actual path metric distribution of node pairs in $T$. In this paper, we study how such path metric distribution inconsistency between $S$ and $T$ affects the accuracy of the network tomography task.

\begin{figure}[tb]
\centering
\includegraphics[width=3.8in]{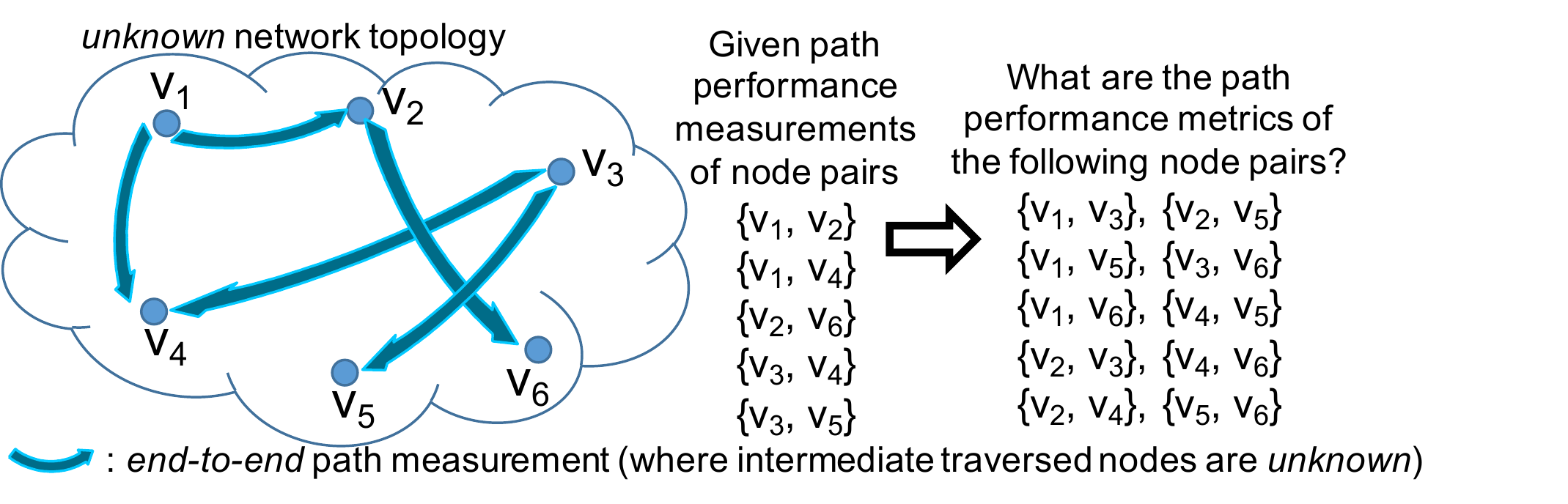}
\caption{Network Tomography in a Sample Network.} \label{Fig:Tomo}
\end{figure}

In sum, as the input, we are only given the basic end-to-end path information for a set of node pairs. 
We then investigate whether such limited information is sufficient to reveal unmeasured network information; see Figure~\ref{Fig:Tomo} for the problem illustration.

\vspace{.5em}
{\bf\emph{Remark:}} Though the measured node pair set $S$ can be any subset of $T$, $S$ needs to cover all nodes in the network. In other words, each node in $V$ appears at least once in a node pair of $S$, i.e., $\bigcup_{\phi \in S} \phi = V$. This is because if there exists node $w\in V$ that is not covered by $S$, then we do not know the existence of $w$; therefore, there is no request on inferring the performance metric of a path starting/terminating at $w$.

\subsection{Objective}
\label{subsec:objective}

For a performance metric of interest (e.g., delay, congestion level, etc.), suppose we are only given the measured end-to-end path performance metrics w.r.t. all node pairs in set $S$. Our first objective is to develop a generic framework to infer the end-to-end path performance metrics for all unmeasured node pairs, i.e., in set $T\setminus S$ (recall that $T$ is the complete node pair set). 
Our second objective is to reconstruct the original network topology based on the given path performance measurements of node pairs in set $S$ when the type of performance metrics allows; see Section~\ref{subsec:TopoReconstruct} for details.

\vspace{.5em}
{\bf\emph{Discussions:}} \textcolor{mycolour}{In practical networks, generally only a small portion of end-to-end path measurements are available.}
Therefore, for the first objective, we aim to establish a framework that exhibits high accuracy when $|S|$ is small, thus applicable to real network monitoring tasks. Note that in cases where some node pairs in $T\setminus S$ are directly connected by links, i.e., neighbors in the underlying \emph{unknown} network topology, their inferred path performance metrics correspond to their link performance metrics if these links are selected for routing. 
While regarding the second objective, the significance of it is that the topological information is critical to many network applications and operations, e.g., traffic engineering, fault localization, etc.

\section{Problem Challenges and Relations to Classical Network Tomography Problems}
\label{sec:problemChallenges}

Our problem formulation in Section~\ref{Sect:ProblemFormulation} is a generalization of classical network tomography problems. In particular, let $\textbf{c}$ denote the path performance metric vector of all node pairs in set $S$. Then the network tomography problem can be described by 
\vspace{-.0em}
\begin{equation}\label{Eq:problem formulation}
    \mathbf{R}\otimes\textbf{w}=\textbf{c},
\vspace{-.0em}
\end{equation}
where $\textbf{w}$ is the performance metric vector of all links in the network with entry $w_i$ denoting the performance metric of link $l_i$, $\mathbf{R}=(R_{i,j})$ is called the \emph{routing matrix} with each entry $R_{i,j}\in \{0,\: 1\}$ representing whether link $l_j$ is present on the path between the $i$-th node pair in $S$, and $\otimes$ is the operator indicating how link metrics are related to the end-to-end path metrics. \textcolor{mycolour}{In particular, the meaning of $\otimes$ depends on the problem considered as described as follows. }

\textcolor{mycolour}{\emph{1) Additive metric tomography:} 
For this class of problems, the common assumption is that $\mathbf{R}$ and $\textbf{c}$ are both known, and $\otimes$ is simply the matrix multiplication. }

 \textcolor{mycolour}{\emph{2) Boolean metric tomography:} For Boolean metric tomography, all performance metrics are binary, where $0$ represents ``normal'' and $1$ represents ``failed''. In this case, $\otimes$ is Boolean matrix product, i.e., $c_{i}=\vee_{j} (R_{i,j}\wedge w_{j})$. }

 \textcolor{mycolour}{\emph{3) Congestion level (or bandwidth) tomography:} 
For congestion level tomography, operator $\otimes$ finds the most congested link in the given path, i.e., $c_{i}=\max_{j} (R_{i,j} w_{j})$. While for bandwidth tomography, $\otimes$ finds the link with the minimum bandwidth, i.e., $c_{i}=\min_{j} (R_{i,j} w_{j})$. }

{\bf\emph{Discussions:}}  \textcolor{mycolour}{For all these classical problems, $\mathbf{R}$ is assumed to be known and the number of faulty links is generally bounded. Then, $\textbf{w}$ and $\mathbf{R'}$, which is the routing matrix corresponding to the unmeasured node pairs in $T\setminus S$, are inferred for computing the path performance metrics for node pairs in $T\setminus S$ via $\mathbf{R'}\otimes\textbf{w}$. In comparison, our problem setting is more relaxed. Without any assumptions on $\textbf{w}$ or $\mathbf{R}$, we are tasked to determine $\mathbf{R'}\otimes\textbf{w}$ directly with the given $\textbf{c}$. This problem is extremely challenging as we are only given $\textbf{c}$ and  the number of entries in $\textbf{w}$ is unknown, since we have no knowledge about the network topology.  We design our network tomography framework to infer $\mathbf{R'}\otimes\textbf{w}$ directly since even  the  metrics $\textbf{w}$ is inferred somehow, it is still \emph{not} sufficient to determine end-to-end path metrics . This is because without knowing the underlying principles that govern the end-to-end routing, $\textbf{w}$ cannot uniquely determine the end-to-end path performance. Hence, we generalize classical tomography issues by removing all assumptions made on $\textbf{w}$ and $\textbf{R}$.}


\begin{figure}[b]
\centering
\includegraphics[width=2.8in]{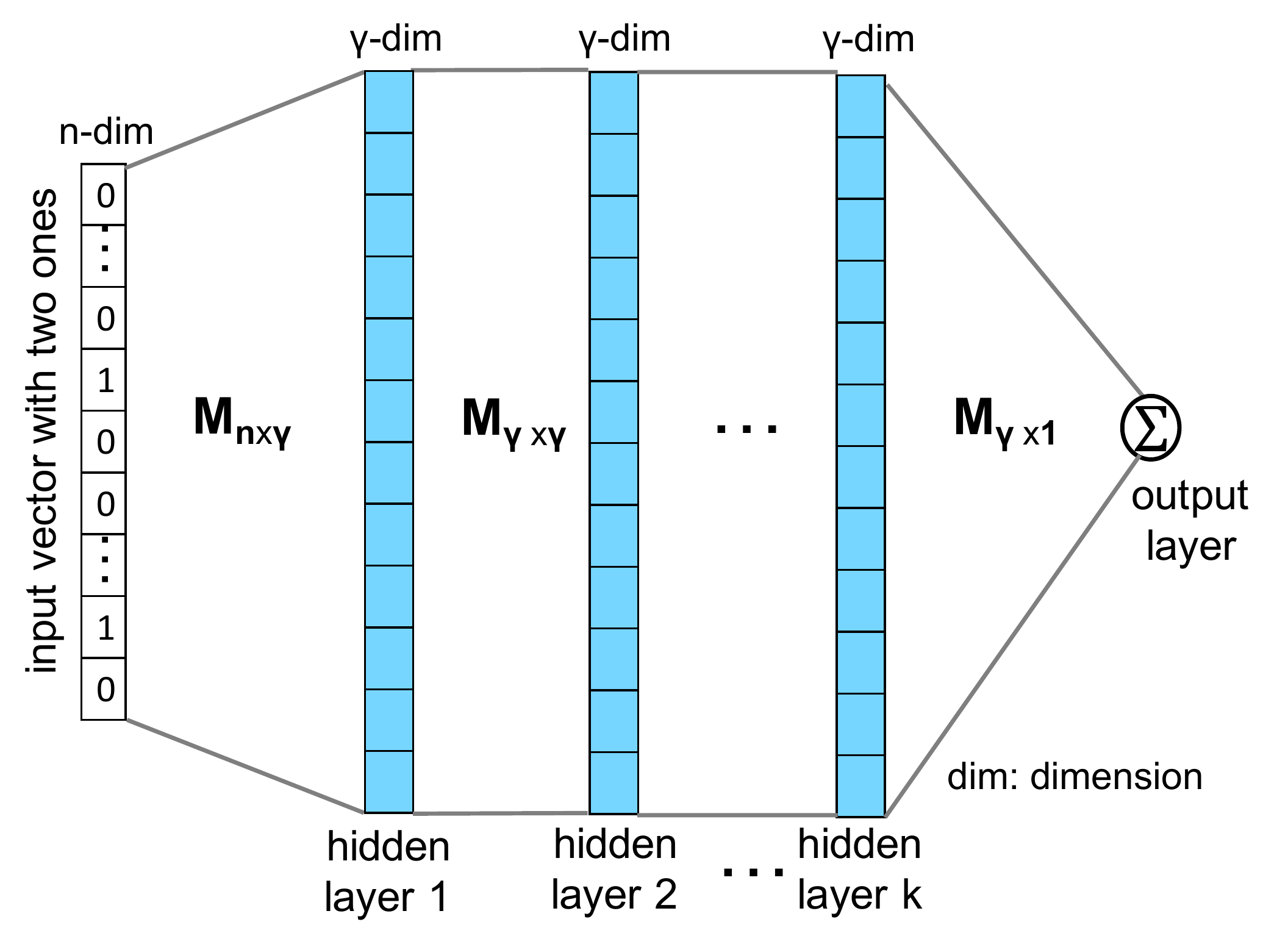}
\caption{Neural-network-based network tomography framework (\emph{NeuTomography}).} \label{Fig:NNModel}
\end{figure}

\section{Generic Network Tomography Framework}\label{sec:NNmodel}

In this section, we propose a neural-network-based network tomography framework to address our objectives. 

\subsection{Neural-Network-Based Tomography Framework (NeuTomography)}
\label{subsec:NNModel}
For neural networks, they are shown to be exceptionally powerful in the field of machine learning \cite{Goodfellow16book}. Moreover, the universal approximation theorem \cite{Hornik91} proves that neural networks are capable of approximating any non-linear functions as long as the hidden layer consists of sufficient number of neurons. As such, we use neural networks to address the potential non-linearality in our network tomography problem.

The neural network is a mathematical architecture where the training variables are continuous. In contrast, both routing matrices $\mathbf{R}$ and $\mathbf{R'}$ in our problem are binary (see the discussions Section~\ref{sec:problemChallenges}). Nevertheless, our ultimate goal is to determine $\mathbf{R'}\otimes\textbf{w}$ rather than individual $\mathbf{R'}$ and $\textbf{w}$. Therefore, we propose to relax the binary values in the routing matrices $\mathbf{R}$ and $\mathbf{R'}$ to continuous values ranging from $0$ to $1$, thus forming \emph{stochastic routing matrices}, denoted by $\widetilde{\mathbf{R}}$ and $\widetilde{\mathbf{R}}'$, respectively. Then entry $\widetilde{R}_{i,j}$ in $\widetilde{\mathbf{R}}$ (or $\widetilde{R}'_{i,j}$ in $\widetilde{\mathbf{R}}'$) indicates the \emph{probability} that link $l_j$ exists on the path connecting the $i$-th node pair in $S$ (or $T\setminus S$). By such routing matrix relaxation, we propose a neural-network-based network tomography framework, called \emph{NeuTomography}, as shown in Figure~\ref{Fig:NNModel}.

Figure~\ref{Fig:NNModel} is a neural network model consisting of $k$ fully-connected hidden layers \cite{Goodfellow16book}, where each hidden layer contains $\gamma$ neurons. Here, $\gamma$ is the estimated number of links in the network. Note that the exact number of links is unknown, and we discuss later how the value of $\gamma$ is estimated and how $\gamma$ affects the inference accuracy. For this model, each input is a node pair from set $S$. At the input layer, the node pair, say $v_1$ and $v_2$, is mapped to an $n$-dimensional ``one-hot'' vector $\mathbf v_0$ (recall that $n$ is the total number of nodes), where only positions corresponding to $v_1$ and $v_2$ are $1$ and $0$ elsewhere.  
Next, as in typical fully connected neural networks, $\mathbf v_0^T$ is multiplied by an $n\times \gamma$ matrix $\mathbf M_1$ and then added by a bias vector $\mathbf b_1$. The resulting $\mathbf v_0^T\mathbf M_1+\mathbf b_1^T$ is passed to hidden layer $1$ and taken as input by an activation function \cite{Goodfellow16book} $\sigma(\cdot)$, i.e., hidden layer $1$ outputs $\sigma(\mathbf v_0^T\mathbf M_1+\mathbf b_1^T)$. Each of the following-up hidden layers has the same activation function $\sigma(\cdot)$ and operates by the same way. Thus, let the output vector from hidden layer $j-1$ be $\mathbf v_{j-1}$; then the output from hidden layer $j$ ($j\leq k$) is $\mathbf v_{j}^T=\sigma(\mathbf v_{j-1}^T\mathbf M_j+\mathbf b_j^T)$, where $\mathbf M_j$ is a $\gamma\times\gamma$ weight matrix between hidden layers $j-1$ and $j$, and $\mathbf b_j$ is the corresponding bias vector. Finally, $\mathbf v_{k}^T$ generated by hidden layer $k$ is multiplied by a $\gamma\times 1$ weight vector $\mathbf m$, i.e., $\mathbf v_{k}^T\mathbf{m}$, as the final path performance metric between the input node pair $v_1$ and $v_2$ at the output layer (only one neuron and \emph{no} activation function or bias in the final output layer).

\vspace{.5em}
{\bf\emph{\textcolor{mycolour}{Design Intuitions of NeuTomography}:}} 
In NeuTomography, we select \emph{sigmoid} \cite{Goodfellow16book} as the activation function $\sigma(\cdot)$ across all hidden layers, i.e., to represent the  probability of each link appearing on paths. 
Its design intuition is as follows: 

\textcolor{mycolour}{\emph{1)} When performance metrics are additive, for the input node pair, say the $i$-th node pair in set $S$, intuitively, the purpose of all hidden layers is to compute the $i$-th row $\widetilde{\mathbf R}_{i,:}$ of the stochastic routing matrix $\widetilde{\mathbf R}$ (each entry value is between $0$ and $1$). 
Then the weight vector $\mathbf m$ connecting to the output layer represents the approximated metrics for all links in the network.
Moreover, the mapping from node pairs to the routing matrix is highly complicated and likely to be non-linear. Therefore, we use multiple ($k$) hidden layers to capture such relations, where each additional layer tries to refine the probability of each link appearing on a particular path.}

\textcolor{mycolour}{\emph{2)} When performance metrics are non-additive, e.g., congestion level and bandwidth, the design intuition is that 
for the $i$-th input node pair, the goal of the $k$-th hidden layer is to output a ``one-hot'' vector $\mathbf v_k$ with $1$ in only one position and $0$ elsewhere. In this ``one-hot'' vector, the position with value $1$ corresponds the most problematic link for the congestion level and bandwidth tomography.
On the other hand, since our objective is to accurately predict the product of $\mathbf v^T_k$ and $\mathbf m$ and $\mathbf v_k$ and $\mathbf m$ are not unique for the same product, 
we only need to train the neural network model for such product instead of individual $\mathbf v_k$ and $\mathbf m$, thus easing the training process.}\looseness=-1

\vspace{.5em}
{\bf\emph{Discussions:}} Our goal is to predict $\mathbf R'\otimes \mathbf w$ by only using $\mathbf R\otimes \mathbf w$, where $\mathbf R$ and $\mathbf R'$ are related by the underlying routing protocol(s), and no prior knowledge about $\mathbf R$, $\mathbf R'$, or $\mathbf w$ is available. The gist of NeuTomography is to capture such relations among $\mathbf R$, $\mathbf R'$, and $\mathbf w$ by the estimated number of links $\gamma$ and multiple hidden layers. In Section~\ref{sec:experiments}, by extensive experiments, we show that by using only a small portion of measurements as the input data, 
NeuTomography is capable of learning the accurate relations between $\mathbf R$ and $\mathbf R'$ irrespective of the type of performance metrics. Moreover, we show that NeuTomography is robust against the estimation error of the number of links ($\gamma$). 

\subsection{Path Measurement Augmentation}

In Section~\ref{subsec:NNModel}, NeuTomography purely utilizes the given measurements of node pairs in $S$ to predict the path performance of node pairs in $T\setminus S$. However, the measured performance metric distribution might be different from the actual performance metric distribution as $S$ can be any subset of $T$; moreover, for the training data, only a small percentage of node pairs are measured (e.g., for the experiments in Section~\ref{sec:experiments}, $|S|/|T|\leq 30\%$), thus potentially causing model overfitting \cite{Goodfellow16book}. For instance, if the input data only include node pairs that are less than $3$-hop away, then the predicted distance for unmeasured node pairs are also up to $3$ hops though the network diameter (which is not directly given in the input data) might be substantially larger than $3$. 
As such, we propose one algorithm that leverages $S$ to construct additional input data to improve the prediction accuracy.

\subsubsection{Motivation and Algorithm Sketch}
For each node pair $\phi\in S$, let $d_\phi$ denote the measured path performance metric w.r.t. $\phi$, \textcolor{mycolour}{and $V$ the set of nodes appearing in the node pairs of set
$S$.} Then, the measurement data can be directly mapped to a weighted graph $\mathcal{G}'=\big(V, S, \{d_\phi\}_{\phi\in S}\big)$, where $V$ is the set of vertices, $S$ specifies the end-points of all edges in $\mathcal{G}'$, and $\{d_\phi\}_{\phi\in S}$ (i.e., path performance metrics w.r.t. $S$ in the input data) are the corresponding edge weights in $\mathcal{G}'$.  In $\mathcal{G}'$, for each node pair $\mu$ in $T\setminus S$, there exists a path $\mathcal{P}_\mu$ connecting the nodes in $\mu$ if $\mathcal{G}'$ is a connected graph. \textcolor{mycolour}{If $\mathcal{G}'$ is disconnected and there exist node pairs which are not connected by any paths in $\mathcal{G}'$, then these node pairs are not selected for augmented input data.} Our idea is to use the performance metric of $\mathcal{P}_\mu$, denoted by $\widetilde{d}_{\mu}$, as the estimation of the real path metric of the unmeasured node pair $\mu$, and feed $(\mu, \widetilde{d}_{\mu})$ to NeuTomography as additional data. Then $\widetilde{d}_{\mu}$ is updated iteratively using its initial estimation and the predicted value by NeuTomography. Lastly, the refined $\widetilde{d}_{\mu}$ is returned as the final inferred path performance metric for $\mu\in T\setminus S$. Based on this idea, we propose a tomography algorithm with augmented data, called Path Augmented Tomography (PAT).

\subsubsection{Path Augmented Tomography (PAT)}
Complete algorithm of PAT is presented in Algorithm~\ref{Alg:NNAugmentation}. In PAT, we first compute the path performance estimation $\widetilde{d}_{\mu}$ for each node pair $\mu$ in $T\setminus S$ by lines~\ref{PAT:estimateS}--\ref{PAT:estimateE}. \textcolor{mycolour}{Path performance estimations are carried out such that $\widetilde{d}_{\mu}$ corresponds to the path with the best performance metric on  $\mathcal{G}'$ w.r.t  the given tomography task.}
Next, with this path performance estimation, we iteratively train the neural network framework. Specifically, from the $|T\setminus S|$ unmeasured node pairs, $\alpha |T\setminus S|$ random node pairs with the estimated path performance values are selected 
(by line~\ref{PAT:selectAdditonal}) and combined with the given measurement data (line~\ref{PAT:combineData}) as the augmented training data to train NeuTomography (line~\ref{PAT:train}).  \textcolor{mycolour}{Note that $\beta$ in line~\ref{PAT:update} equals zero for node pairs that are not within the same component when $\mathcal{G}'$ is disconnected.} After this training process, the path performance estimation $\{\widetilde{d}_{\mu}\}_{\mu \in T\setminus S}$ is updated by lines~\ref{PAT:updateS}--\ref{PAT:updateE}. In particular, parameter $\beta$ ($0\leq\beta<1$) is employed in line~\ref{PAT:update} to balance the estimated and the predicted value so as to avoid overfitting. Such training and value updating process is repeated until the maximum number of iterations is reached. Finally, Algorithm~\ref{Alg:NNAugmentation} outputs $\{\widetilde{d}_{\mu}\}_{\mu \in T\setminus S}$.

\vspace{.5em}
\noindent{\bf\emph{Discussions of Algorithm~\ref{Alg:NNAugmentation}:}}
\textcolor{mycolour}{There are two key operations in Algorithm~\ref{Alg:NNAugmentation}, i.e., the first  ``foreach'' loop which initializes path performance metric estimations for unmeasured node pairs based on $\mathcal{G}'$; and the second ``while'' loop which iteratively updates these estimations using predictions made by NeuTomography while it is being trained.  The estimation update process is regulated by the parameter $\beta$ which controls the amount of new estimations allowed. Such soft update process ensures the stability of the training process of NeuTomography and can be found in some existing works on machine learning. Intuitively, as the predictions made by NeuTomography become more accurate while it is being trained, the higher accuracy of the predicted data in turn helps the training process. }
In Section~\ref{sec:experiments}, we conduct extensive experiments to understand how such path augmented tomography and tunable parameters ($\alpha$, $\beta$, and \#iterations) affect the performance inference accuracy. 


\begin{algorithm}[tb]
\SetKwInOut{Input}{input}\SetKwInOut{Output}{output}
\Input{Path performance measurements for each node pair in $S$: 
$\{(\phi,d_\phi)\}_{\phi\in S}$, proportion of additional data $\alpha$ ($0<\alpha<1$), update weight $\beta$ ($0\leq\beta<1$)}
\Output{Inferred path performance metric for each node pair in $T\setminus S$}
\BlankLine
\ForEach{node pair $\mu$ in $T\setminus S$\label{PAT:estimateS}}
    {$\widetilde{d}_\mu\leftarrow$ path performance estimation for $\mu$ using the input measurement data $\{(\phi,d_\phi)\}_{\phi\in S}$\;\label{PAT:estimate}
    }\label{PAT:estimateE}
\While{the maximum number of iterations is not reached\label{PAT:nIterations}}
    {
    randomly choose $\alpha|T\setminus S|$ node pairs from $T\setminus S$ to form the augmented node pair set $A$\;\label{PAT:selectAdditonal}
    $training\_data=\{(\phi,d_\phi)\}_{\phi\in S}\cup \{(\theta,\widetilde{d}_\theta)\}_{\theta\in A}$\;\label{PAT:combineData}
    use the above $training\_data$ to train NeuTomography\;\label{PAT:train}
    \ForEach{$\mu\in T\setminus S$\label{PAT:updateS}}
        {
        $\widetilde{d}_\mu\leftarrow\beta\cdot\widetilde{d}_\mu + (1-\beta)\cdot NT(\mu)$\tcp*{$NT(\mu)$: the path performance metric for $\mu$ predicted by NeuTomography after training in line~\ref{PAT:train}}\label{PAT:update}
        }\label{PAT:updateE}
    }
return $\widetilde{d}_\mu$ for each $\mu\in T\setminus S$\;
\caption{Path Augmented Tomography (PAT)}
\label{Alg:NNAugmentation}
\end{algorithm}
\normalsize

\begin{figure}[tb]
\centering
\includegraphics[width=3.3in]{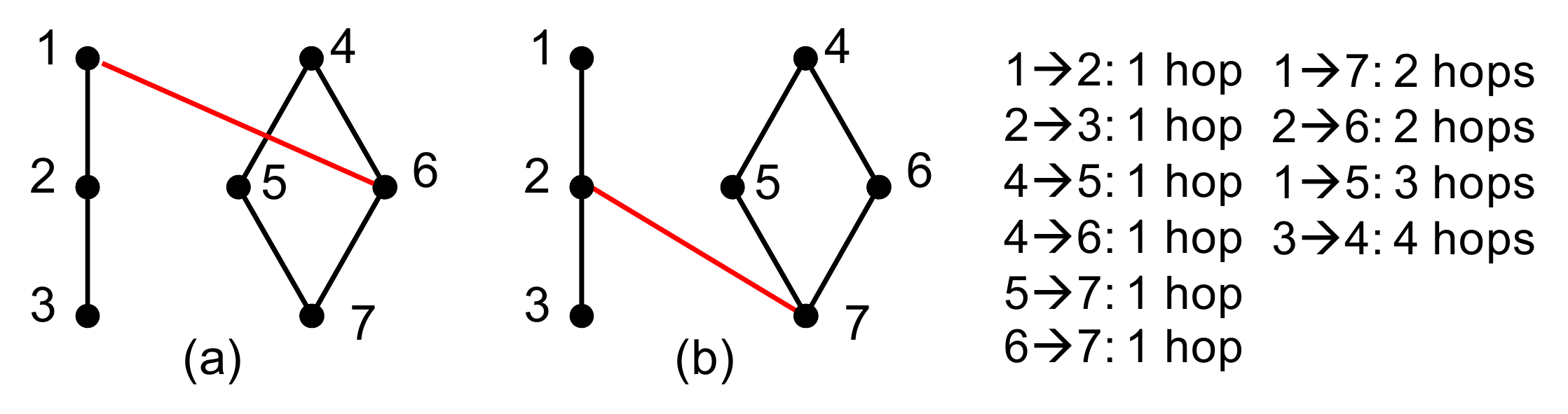}
\caption{Non-uniqueness of network topologies corresponding to the same set of path measurements.} \label{Fig:TopologyReconstruction}
\end{figure}

\subsection{Network Topology Reconstruction}
\label{subsec:TopoReconstruct}
Our network topology reconstruction task is more challenging than prior traceroute-based router-level topology inference works \cite{Spring02SIGCOMM,Gunes07IMC,Mao03SIGCOMM,Eriksson08SIGCOMM,Eriksson11INFOCOM,Yao03INFOCOM}. This is because traceroute (may be blocked in real networks for security reasons) enables a hop-by-hop approach, where each measurement path contains the per-hop information of all traversed nodes; such information, by contrast, is not available in our input data. 

Given that only the path measurements for set $S$ are available, among all the performance metrics of interest, only the minimum number of hops between two nodes provides useful information on the underlying network topology. This is because, for other path performance metrics, e.g., delay and congestion level, even if there exists a link between two nodes, such link may \emph{not} be used in any communication paths due to its poor performance or routing constraints. Therefore, the link performance metric between two neighboring nodes can be arbitrarily larger or smaller than their path performance metric. Thus, for network topology reconstruction, in this section, we only focus on the case where the path performance metric is the number of hops.

However, even the path measurements w.r.t. set $S$ are the number of hops, the corresponding network topology is not unique. For instance, in Figure~\ref{Fig:TopologyReconstruction}, given the same measurements, both topologies in Figure~\ref{Fig:TopologyReconstruction} suffice and exist with equal probabilities. 
In this regard, we introduce the following definition to extend the description of network topologies.

\begin{table}[t]
\renewcommand{\arraystretch}{1.3}
\caption{\textcolor{mycolour}{AS Topologies in Rocketfuel and ITDK}} \label{t ISP}
\centering
\tabcolsep=0.14cm
\renewcommand{\arraystretch}{0.9}
\begin{tabular}{c|c|>{\color{mycolour}}c|>{\color{mycolour}}c|c|c|c|c|c}
\hline
\multirow{2}{*}{AS} &  \multirow{2}{*}{\thead{\# \\ nodes}}&  \multirow{2}{*}{D} &  \multirow{2}{*}{ASPL} & \multirow{2}{*}{\thead{\emph{unknown} \\ \#links}}  & \multicolumn{2}{c|}{\thead{\emph{unknown} \\ link weights}} & \multicolumn{2}{c}{\thead{\emph{unknown} link\\ congestion levels}}\\
\cline{6-9}		
 &    &   &  &  & mean & var & mean & var\\
\hline							
3967 		&	201	& 11 & 4.8 &	434	& 4.2 & 14.5 & 5.2 & 78.0 \\
\hline							
3257 	&	240	& 14 & 5.1&	404	& 4.8 & 18.7 & 4.3 & 41.3\\
\hline							
1221 	&	318	& 13 & 5.0&	758	& 2.0 & 1.1 & 2.8 & 9.3\\
\hline
15706	&	325	&  7  &  3.17	&	874	        &  2.0 & 1.1  &  2.8 & 9.3\\
\hline
\end{tabular}
\end{table}
\normalsize

\begin{definition}
\label{def:extA}
For network $\mathcal{G}$, $\mathbf A^{(m)}$ ($m$ is an integer) is the \emph{$m$-extended adjacency matrix} of $\mathcal{G}$, where $A^{(m)}_{i,j}=1$ if nodes $i$ and $j$ are $m$-hop away in $\mathcal{G}$ and $A^{(m)}_{i,j}=0$ otherwise.
\end{definition}

When $m=1$, the extended adjacency matrix is reduced to a regular adjacency matrix. Comparing $\mathbf A^{(m)}$ ($m>1$) to $\mathbf A^{(1)}$, $\mathbf A^{(m)}$ establishes a coarse-grained representation of the network topology. 
Specifically, since $\mathbf A^{(1)}$ is generally not unique (see Figure~\ref{Fig:TopologyReconstruction}), if $\mathbf A^{(m)}$ is accurate when $m$ is small, then the collection of the constructed extended adjacency matrices $\{\mathbf A^{(1)},\mathbf A^{(2)},\ldots,\mathbf A^{(m)}\}$ jointly determine the network topology with various granularity and accuracy tradeoffs, through which the network operators can perform network planning tasks based on the optimization objectives and the topology granularity/accuracy preference levels. In Section~\ref{sec:experiments}, we show that although $\mathbf A^{(1)}$ is likely to be inaccurate, $\mathbf A^{(2)}$ already achieves high accuracy for some networks. 

Thus, with Definition~\ref{def:extA}, we reconstruct the $m$-extended adjacency matrix for different $m$. 
The reconstruction algorithm for approximating $\mathbf A^{(m)}$ is simple: For each node pair $\{i,j\}\in T$, if the (inferred or given) performance metric of the path between $i$ and $j$ falls into the region of $(m-0.5,m+0.5]$, then $A^{(m)}_{i,j}=1$; otherwise, $A^{(m)}_{i,j}=0$.


\section{Experiments}
\label{sec:experiments}

\subsection{Input Data}
\label{subsec:InputData}

We evaluate NeuTomography through extensive experiments on \emph{Autonomous System} (AS) networks from both the Rocketfuel \cite{RocketFuel} and ITDK \cite{CAIDA} projects, which represent IP-level connections between backbone/gateway routers of several ASes from major \emph{Internet Service Providers (ISPs)} around the globe. The parameters of selected networks obtained from these two projects are listed in Table~\ref{t ISP}, where AS15706 is from ITDK and others are from Rocketfuel and the last five columns are unknown to NeuTomography (as discussed in Section~\ref{Sect:ProblemFormulation}). 
\textcolor{mycolour}{Note that ``D'' and ``ASPL'' in Table~\ref{t ISP} stand for network diameter and average shortest path length, respectively; both in terms of the number of  hops.} Since the Rocketfuel and ITDK projects do not directly provide path measurements, we consider the following three aspects to generate  path measurements using the available network data for evaluating NeuTomography. 

\vspace{.5em}
{\bf\emph{Remark:}} The purpose here is only to provide a method to generate measurement paths to evaluate NeuTomography. Besides the end-to-end path metrics of selected node pairs, NeuTomography does \emph{not} know anything about link metrics, network topologies, routing strategies, or sampling methods that are used to generate data as discussed below.
\vspace{.5em}



\emph{1) Link metrics.}
Unlike Rocketfuel, ITDK in \cite{CAIDA} does not provide the link metric information; therefore, for the experiment purpose, regarding AS15706, its link metric distribution is approximated by AS1221 in Rocketfuel (which has the similar number of nodes). 
Furthermore, besides these link metric information in Table~\ref{t ISP}, to extensively study NeuTomography, we also consider two other types of link metrics: (i) unweighted link metrics, where there is no link metric, 
and (ii) uniform link metrics, where link metrics in the network are uniformly distributed between $1$ and $10$. 

\emph{2) Routing strategies.} To construct a path between two nodes, two routing strategies are employed: (i) \emph{Min-Hop Routing (MHR)}, where a path incurring the minimum number of hops is selected, and (ii) \emph{Best Performance Routing (BPR)}, where w.r.t. a given performance metric of interest, the path with the best performance metric is selected, e.g., shortest path for the metric of delay, least congested path for the metric of congestion level. 

\emph{3) Sampling methods.} When the above \emph{1)} and \emph{2)} are known, the end-to-end path performance metrics can be obtained for all node pairs (in set $T$). We then sample a subset $S$ of $T$ to form the input data. 
We first consider \emph{random sampling}, where $S$ is randomly picked from $T$. 
Since there may exist constraints on measurable pairs in real networks, we next consider an alternative method, called \emph{monitor-based sampling}. In monitor-based sampling, we first randomly select $\rho$ nodes as monitors; then each monitor pings all other nodes (both monitors and non-monitors) to measure the end-to-end path performance between them. Thus, under monitor-based sampling, 
each node pair in $S$ contains at least one monitor. 
For each of these sampling methods, the sampling ratio $|S|/|T|$, is selected from $\{20\%, 25\%, 30\%\}$ (the number of monitors $\rho$ under monitor-based sampling is tuned such that the required $|S|/|T|$ is reached). 
All path metrics associated with node pairs in $T\setminus S$ serve as the testing data. 

\subsection{Benchmark Solutions}
To study the performance of NeuTomography, it is compared against the following benchmarks.

\emph{1) Minimum Monitor Placement and Determination of All Identifiable Links (MMP+DAIL)}. MMP+DAIL \cite{Ma13IMC,Ma14INFOCOM} is a state-of-the-art tomographic solution for additive performance metrics, under the assumption of \emph{known} network topology and controllable cycle-free routing. In particular, MMP places the minimum number of measuring vantages to ensure all measurement paths are sufficient to accurately compute all link metrics. While DAIL determines all links whose performance metrics are accurately inferable under the given measuring vantages. 
To employ MMP+DAIL as a benchmark, we test it under erroneous topological information as follows. Let $\mathcal{G}=(V,E)$ be the actual network topology ($V$/$E$ set of nodes/links in $\mathcal{G}$) and $\mathcal{G}'=(V',E')$ the perceived network topology by MMP with the topological information error being $\epsilon$ ($0<\epsilon\ll 1$), where $V=V'$ and $E\neq E'$. Specifically, for link $e\in E$, $e\in E'$ with probability $1-\epsilon$; for link $e\notin E$, $e\in E'$ with probability $\epsilon$. Based on the measuring vantages placed by MMP in $\mathcal{G}'$, DAIL determines all link metrics. However, links in $E\setminus E'$ are not visible to DAIL and links in $E'\setminus E$ do not exist. Therefore, DAIL only utilizes links in $E\cap E'$ to construct measurement paths for determining link metrics in $E'$. For edges in $E'$ whose performance metrics cannot be uniquely determined, they are assigned arbitrary values according to the distribution of the inferable link metrics. With these link metrics, if the underlying routing mechanism is given, then the path performance metric for any node pair is computable.




\textcolor{mycolour}{\emph{2) Arbitrary-valued Non-additive Metric Identification (ANMI)}. For non-additive performance metrics, e.g., congestions, most existing tomography approaches \cite{Ma14IMC,Ma17ToN,Bartolini17INFOCOM,Ma15Performance,He16ICDCS,He17DistributedLA,Tootaghaj18Performance,Galesi18ICDCS} target to localize the problematic links under the assumption of known network topology.  Currently, the state-of-the-art approaches are capable of either uniquely locate up to $k$ binary link metrics (normal/failed) \cite{Ma15Performance} or locating only one problematic link and determining its arbitrary-valued link metric \cite{Zarifzadeh12IMC}. To the best of our knowledge, there is no tomographic approach that is capable of handling arbitrary-valued non-additive performance metrics without the constraint of the number of problematic links. In this regard, we employ an artificial method called Arbitrary-valued Non-additive Metric Identification (ANMI) that is similar to range tomography, but without the constraint of the number of problematic links. Specifically, given a tunable threshold parameter $\tau$, when a link performance metric is less than $\tau$, then this link is regarded as normal, and problematic otherwise.  Suppose there exists a method that can uniquely localize all problematic links in the network when the network topology is known. Then assuming we are also given the precise performance metric distributions of normal and problematic links in the network, we further estimate the fine-grained link metrics by generating the estimated values according to these given metric distributions. Finally, similar to MMP+DAIL, the path performance metric for any node pair is computable if the underlying routing mechanism is given.} 

\textcolor{mycolour}{In addition to MMP+DAIL and ANMI that are proposed specifically for network tomography, we also compare NeuTomography against two solutions established in other related areas, which are described in the following. }

\begin{figure*}[tb]
\centering
\subfigure[random sampling (additive)]{\includegraphics[scale=0.28]{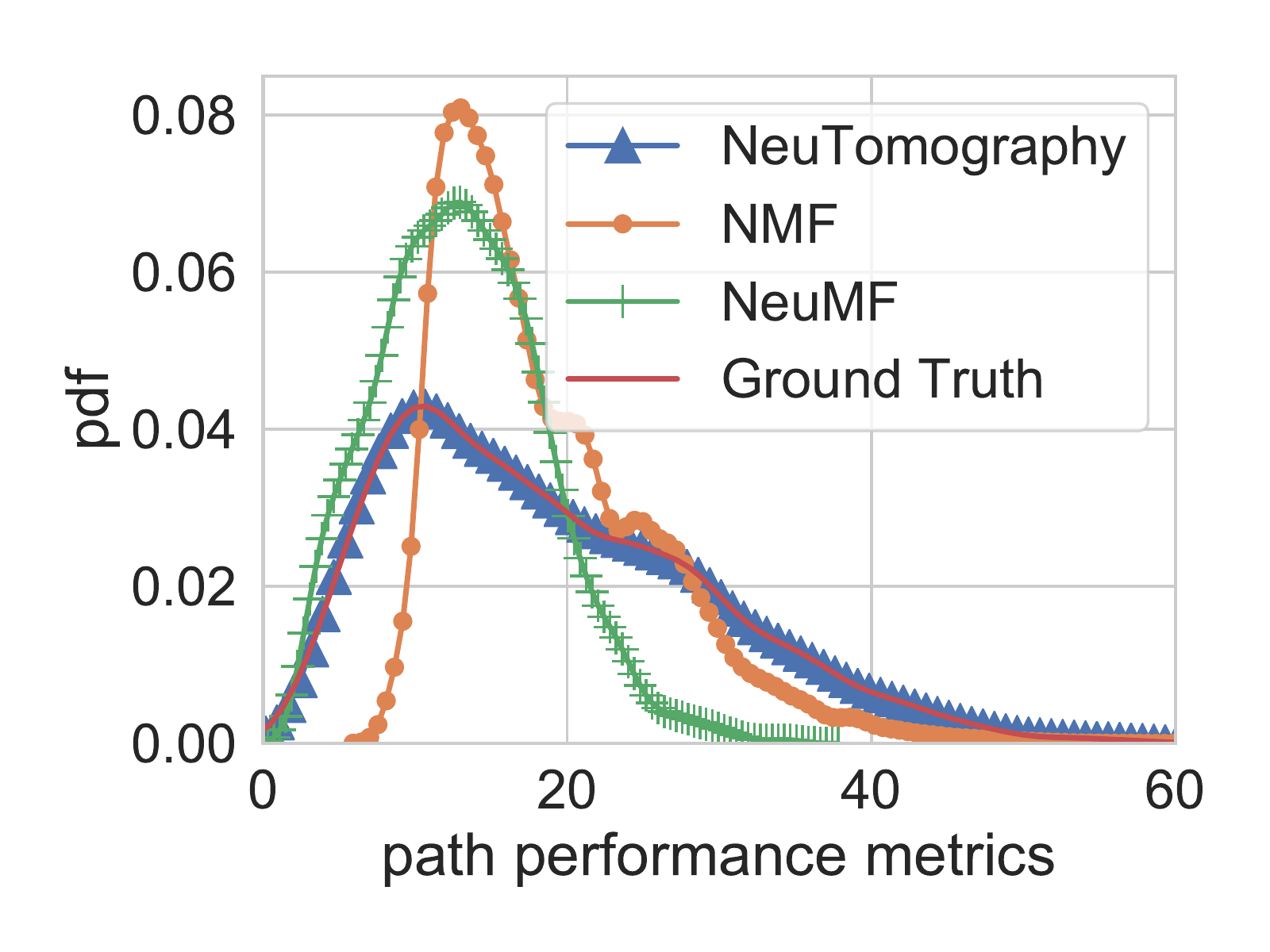}}
\hspace{-1.25em}
\subfigure[monitor-based sampling (additive)]{\includegraphics[scale=0.28]{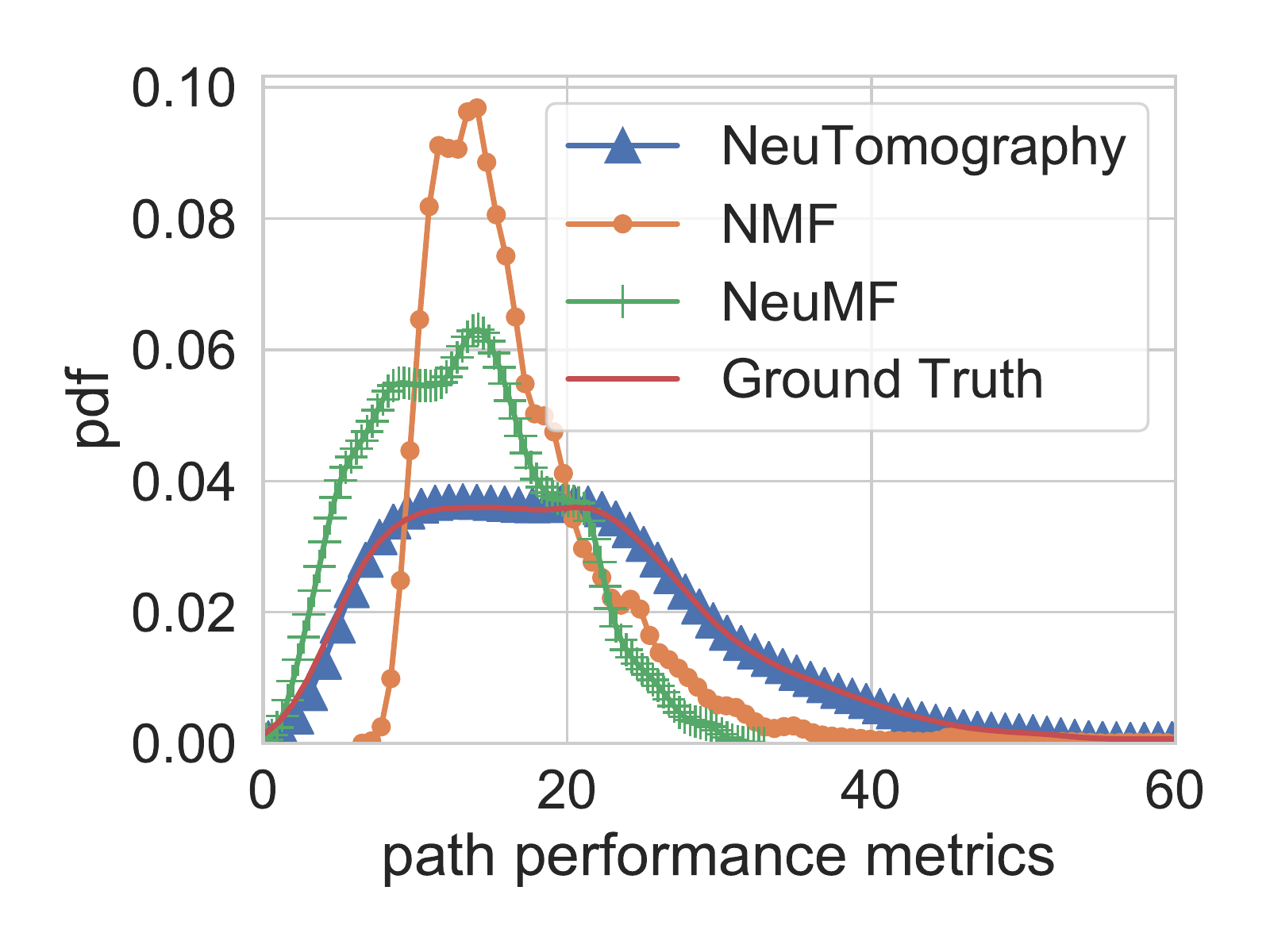}}
\subfigure[random sampling (non-additive)]{\includegraphics[scale=0.28]{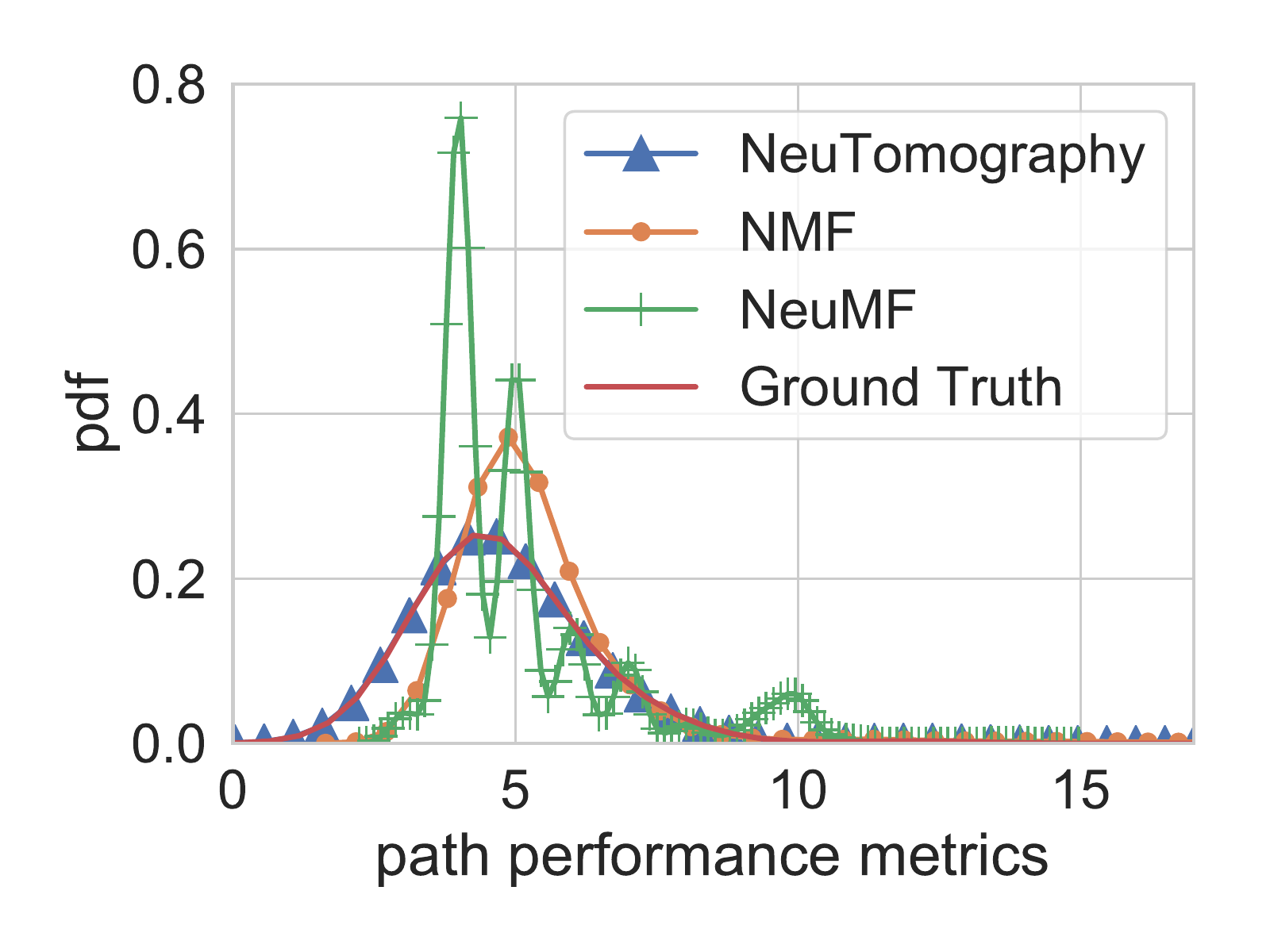}}
\hspace{-1.25em}
\subfigure[monitor-based sampling (non-additive)]{\includegraphics[scale=0.28]{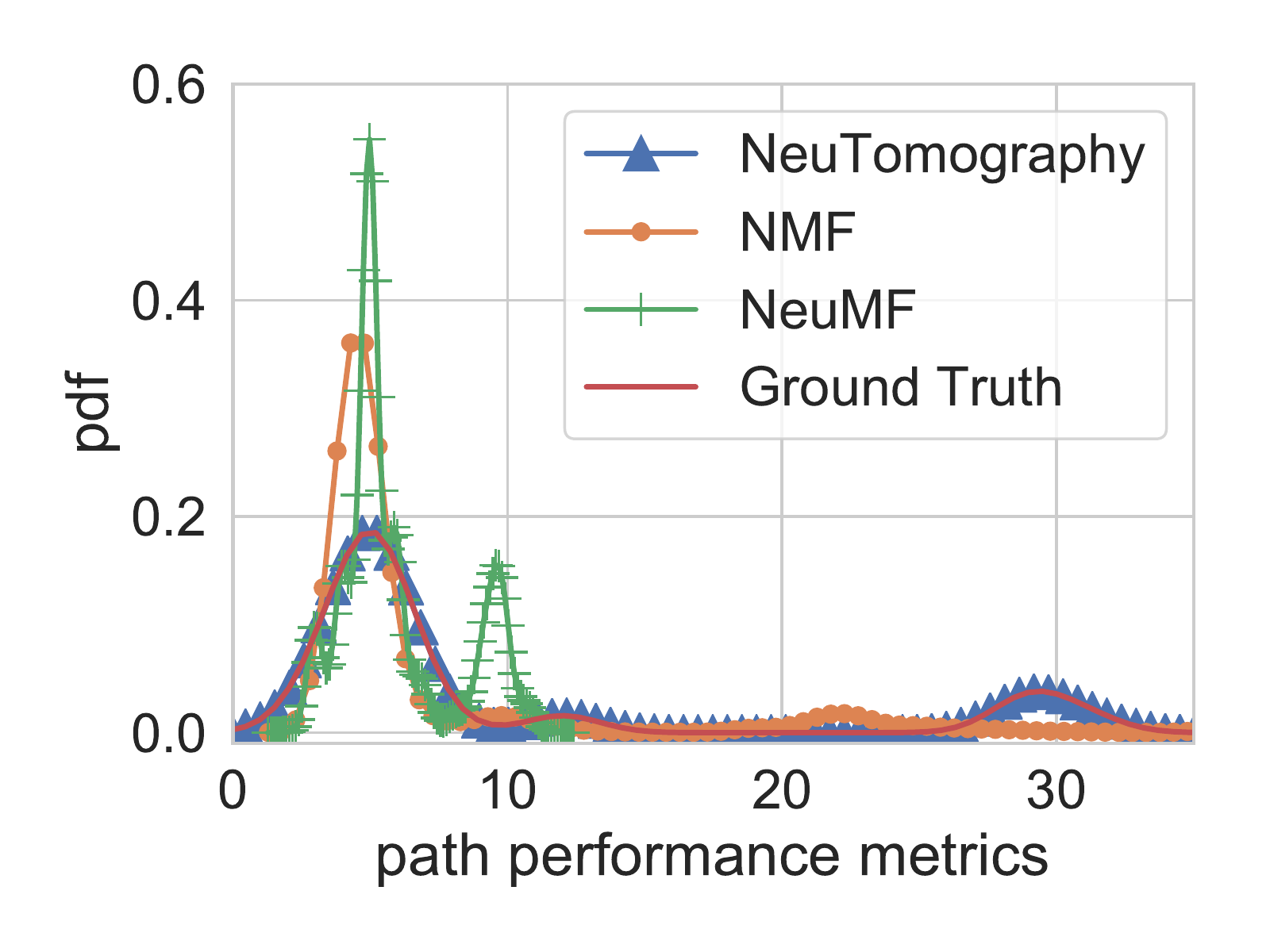}}
\caption{Distribution of the predicted (additive/non-additive) path performance metrics (AS3257 under $|S|/|T|=30\%$, link weights from real data, and best performance routing).}
\label{Fig:MetricDistribution}
\end{figure*}

\emph{3) Non-negative Matrix Factorization (NMF) \cite{Lee00NIPS}}. 
NMF is widely used in recommendation systems, where the goal is to complete the non-negative user-item rating matrix via the product of two lower dimensional matrices. At the high level, recommendation systems and network tomography share similar objectives, as they both target to predict the unknown non-negative entries in a matrix using some given entry values. 
As such, we use NMF as one benchmark solution. 

\emph{4) Neural Matrix Factorization (NeuMF) \cite{He17WWW}}. NeuMF is a neural-network-based solution employing both neural collaborative filtering \cite{Marlin03NIPS} and matrix factorization for recommendation systems. 
To use NeuMF as a benchmark, we tune our measurement data to adapt to the input format of NeuMF, which requires the user-item rating be between $0$ (dislike) and $1$ (like). For non-preferable large performance metrics, e.g., delay and the number of hops, we use the reciprocal of the measured path metrics as the output of NeuMF; for preferable large performance metrics, e.g., bandwidth and delivery ratio, we normalize the path metrics as the NeuMF output. In this way, path metrics ($\geq 1$) with superior (or poor) performance are mapped to values close to $1$ (or $0$). 

\vspace{.5em}
{\bf\emph{Remark:}} \textcolor{mycolour}{ There are \emph{no} existing tomographic solutions that operate under our highly relaxed problem settings. In this regard, we choose MMP+DAIL and ANMI, which have been adapted to our simulation settings, as a representative solutions for traditional tomography problems under strong assumptions. Moreover, NeuTomography is compared to NMF and NeuMF algorithms which demonstrates that although the latter two work well in tasks sharing similar objectives as our tomography tasks, NeuTomography outperforms them due to its customized designs.}
\vspace{.5em}

\subsection{Experiment Settings}

\subsubsection{Input Data for Training} 
We evaluate NeuTomography on three path performance metrics: (i) the number of hops, (ii) \textcolor{mycolour}{accumulated path delays,} and (iii) the path congestion level. Note that the path bandwidth and the binary normal/failed metric are similar to the path congestion level as all of them are determined by the worst performed link; 
therefore, we use the path congestion level as a representative non-additive metric in this section. For each path metric, we generate the input training data via the combination of link metric types, routing strategies, sampling methods, and sampling ratios as discussed in Section~\ref{subsec:InputData}.

\subsubsection{Framework Parameters} 
For NeuTomography, we select the following parameters. As discussed in Section~\ref{subsec:problemInput}, the given measurement data already covers all nodes in the network; therefore, the dimension of the input layer, $n$, in Figure~\ref{Fig:NNModel} is determined. For the number of links $\gamma$ (i.e., the number of neurons in hidden layers), we estimate it by the average node degree (defined as $2\gamma/n$) in real networks. As shown in \cite{barabasi2016network}, in real communication networks, the average node degree is generally between $1$ and $5$. In this regard, we set $\gamma$ as $\gamma=2.5n$. Such $\gamma$ is an overestimation for AS3967, AS3257, and AS1221, but an underestimation for AS15706 (see Table~\ref{t ISP}). In Section~\ref{subsec:pathMetricAccuracy}, we study how such estimation inaccuracy affects the performance. For the number of hidden layers $k$, to balance the accuracy and the training time, we set $k=2$. Furthermore, we select the mean square error (MSE) as the loss function, \emph{Adam} \cite{Adam} (statistical gradient descent based method) as the optimizer, and $1000$ epochs for training. In addition, when the enhanced algorithm PAT is employed, we set $\alpha=15\%$, $\beta=0.6$,  and \#iterations=6 (line~\ref{PAT:nIterations} in Algorithm~\ref{Alg:NNAugmentation}).

\subsection{Path Metric Prediction Accuracy}
\label{subsec:pathMetricAccuracy}

To show the advantages of NeuTomography, we first illustrate the distribution of the predicted performance metrics. For instance, in Figure~\ref{Fig:MetricDistribution}, the predicted metric distribution by NeuTomography almost overlaps with the actual distribution for different performance metric types and sampling methods (PAT is used for monitor-based sampling), while the results by NMF and NeuMF deviate from the actual distribution significantly. Note that 
since MMP+DAIL requires the network topology as input, it cannot be compared with other solutions under the same settings, thus omitted.
In Figure~\ref{Fig:MetricDistribution},
the path performance metrics in the input training data have different distributions under different sampling methods; nevertheless, NeuTomography can recover the actual performance metric distribution for all unmeasured node pairs.

In addition to the predicted performance metric distribution, more importantly, we need to evaluate the path metric prediction accuracy for each unmeasured node pair. As such, in this section, we focus on the mean absolute percentage error (MAPE) as the accuracy evaluation metric; the corresponding results under different experiment settings, i.e., (i) link metrics that are \emph{un}weighted (UN), from \emph{re}al data (Re), or \emph{u}niformly \emph{d}istributed (UD), (ii) best performance routing (BPR) or min-hop routing (MHR), and (iii) random or monitor-based sampling, are reported in Tables~\ref{t_add}--\ref{t_congNeuMF}. 
In addition, we also repeat each experiment under $\gamma=2n$ and $\gamma=3n$, and get similar results, thus omitted for page limitations. These results therefore confirm the robustness of NeuTomography against link number estimation errors.

\begin{table*}[t]
	\renewcommand{\arraystretch}{1.3}
	\caption{Path Performance Prediction Error (MAPE in \%) of NeuTomography for Additive Metrics (UN/Re/UD: link metrics that are unweighted/from real data/uniformly distributed, BPR: best performance routing, MHR: min-hop routing, monitor: monitor-based sampling, m+PAT: monitor-based sampling input data and PAT is applied)} \label{t_add}
	\centering
	\tabcolsep=0.11cm
	\renewcommand{\arraystretch}{0.9}
	\begin{tabular}{c|c|c|c|c|c|c|c|c|c|c|c|c|c|c|c|c|c|c|c|c|c}	
		\hline
		\multicolumn{2}{c|}{ }  & \multicolumn{5}{c|}{AS3967} &  \multicolumn{5}{c|}{AS3257} &  \multicolumn{5}{c|}{AS1221} &  \multicolumn{5}{c}{AS15706}\\		\hline
		\multirow{2}{*}{$\frac{|S|}{|T|}$} & sampling & \multicolumn{3}{c|}{BPR} &  \multicolumn{2}{c|}{MHR} & \multicolumn{3}{c|}{BPR} &  \multicolumn{2}{c|}{MHR} & \multicolumn{3}{c|}{BPR} &  \multicolumn{2}{c|}{MHR}   & \multicolumn{3}{c|}{BPR} &  \multicolumn{2}{c}{MHR}\\
		\cline{3-22}
		& method & UN & Re & UD & Re & UD & UN &  Re & UD  & Re & UD &  UN & Re & UD & Re & UD & UN & Re & UD & Re & UD\\
		\hline					
		\multirow{3}{*}{20\%}	&	random	&	\bf{8.8}  &	\bf{7.3}  &	\bf{7.7}  &	36.0  &	22.4 &	\bf{7.9}  &	\bf{5.9}  &	\bf{6.0}  &	18.0  &	\bf{14.6}  &	\bf{2.9}  &	\bf{3.4}  &	\bf{3.0}  &	\bf{6.7} &	\bf{6.3}  &	\bf{1.3}  &	\bf{1.8}  &	\bf{1.5}  &	\bf{2.3}  &	\bf{3.1}\\
		&	monitor	&	40.4  &	36.4  &	41.9  &	75.9  &	46.7  &	36.2  &	28.5  &	26.8  &	54.8  &	45.1   &	45.1  &	24.3  &	19.3  &	41.4  &	51.1 &	26.7  &	\bf{14.7}  &	\bf{8.4}  &	\bf{11.4}  &	39.6  \\
		&	m+PAT	&	23.3  &	22.4  &	21.4  &	34.2  &	32.9 & 23.8  &	\bf{12.1}  &	15.4  &	28.6  &	21.4	&	26.6  &	\bf{11.1}  &	\bf{9.7}  &	17.0  &	22.7  &		23.6  &	\bf{8.3}  &	\bf{7.0}  &	\bf{9.9}  &	\bf{12.3}  \\
		\hline
		
		\multirow{3}{*}{25\%}	&	random	&	\bf{8.1}  &	\bf{8.0}  &	\bf{7.4}  &	28.3  &	15.9 &	\bf{6.7}  &	\bf{5.4}  &	\bf{5.7}  &	15.4  &	\bf{10.0} &	\bf{2.0}  &	\bf{3.0}  &	\bf{2.2}  &	\bf{5.1}  &	\bf{6.5}  &	\bf{0.9}  &	\bf{1.4} &	\bf{1.4}  &	\bf{2.1}  &	\bf{2.6}  \\
		&	monitor	&	35.8  &	36.9  &	39.5  &	75.8  &	41.6  &	32.9  &	16.9  &	25.4  &	39.3  &	42.0  &	42.0  &	19.5  &	17.3  &	34.5  &	40.2  &	15.2  &	\bf{8.1}  &	\bf{7.3}  &	\bf{10.2}  &	\bf{9.0}  \\
		&	m+PAT	& 20.3  &	\bf{14.6}  &	17.4  &	32.8  &	28.0	&18.1  &	\bf{11.5}  &	\bf{14.3}  &	26.1  &	20.8	&	18.7  &	\bf{9.2}  &	\bf{9.2}  &	15.4  &	21.5  &	\bf{14.9}  &	\bf{5.3}  &	\bf{4.3}  &	\bf{9}  &	\bf{6.8}  \\
		\hline
		
		\multirow{3}{*}{30\%}	&	random	&	\bf{6.5}  &	\bf{5.7}  &	\bf{6.6}  &	22.2  &	\bf{14.4}  &	\bf{5.9}  &	\bf{4.0}  &	\bf{4.5}  &	\bf{13.2}  &	\bf{10.0}  &	\bf{2.4}  &	\bf{2.6}  &	\bf{2.3}  &	\bf{5.2}  &	\bf{5.2}  &	\bf{0.6}  &	\bf{1.2}  &	\bf{1.3}  &	\bf{2.1}  &	\bf{2.5}\\
		&	monitor	&	32.7  &	31.0  &	18.2  &	49.2  &	40.6   &	\bf{9.1}  &	\bf{14.5}  &	21.0  &	28.5  &	40.6   &	35.8  &	17.1  &	\bf{14.0}  &	25.1  &	39.0   &	\bf{13.3}  &	\bf{8.1}  &	\bf{3.0}  &	\bf{3.9}  &	\bf{7.9}  \\
		&	m+PAT	&18.7  &	\bf{14.2}  &	\bf{12.4}  &	33.7  &	28.1	&	\bf{9.1}  &	\bf{8.6}  &	\bf{13.2} &	25.7  &	19.5	&	18.4  &	\bf{7.2}  &	\bf{6.8}  &	\bf{14.9}  &	19.9   &	\bf{13.0}  &	\bf{5.3}  &	\bf{3.0}  &	\bf{3.8}  &	\bf{6.8}  \\
		\hline						
	\end{tabular}
\end{table*}
\normalsize

\begin{table*}[t]
\caption{Path Performance Prediction Error (MAPE in \%) of MMP+DAIL for Additive Metrics} \label{t_MMPDAIL}
	\centering
	\tabcolsep=0.14cm
	\renewcommand{\arraystretch}{0.9}
	\begin{tabular}{c|c|c|c|c|c|c|c|c|c|c|c|c|c|c|c|c|c|c|c|c}	
		\hline
		 & \multicolumn{5}{c|}{AS3967} &  \multicolumn{5}{c|}{AS3257} &  \multicolumn{5}{c|}{AS1221} &  \multicolumn{5}{c}{AS15706}\\	
		
		\hline
		topology & \multicolumn{3}{c|}{BPR} &  \multicolumn{2}{c|}{MHR} & \multicolumn{3}{c|}{BPR} &  \multicolumn{2}{c|}{MHR} & \multicolumn{3}{c|}{BPR} &  \multicolumn{2}{c|}{MHR}   & \multicolumn{3}{c|}{BPR} &  \multicolumn{2}{c}{MHR}\\
		\cline{2-21}
		 error & UN & Re & UD & Re & UD & UN &  Re & UD  & Re & UD &  UN & Re & UD & Re & UD & UN & Re & UD & Re & UD\\
		\hline
		0.5\%	&	21.1	&	23.7	&	29.0	&	36.7	&	30.6	&	21.9	&	35.9	&	32.1	&	35.8	&	28.8	&	22.7	&	30.4	&	29.6	&	30.9	&	28.4	&	\bf6.3	&	20.6	&	29.6	&	22.6	&	24.3	\\
1\%	&	27.1	&	32.5	&	34.2	&	41.9	&	33.5	&	32.8	&	44	&	50.5	&	49.8	&	43.3	&	30.6	&	41.2	&	43.8	&	40.4	&	39.7	&	\bf9.9	&	30.3	&	39.4	&	33.1	&	37.5	\\
2\%	&	35.3	&	44.1	&	47.4	&	47.6	&	43.0	&	39.9	&	54.1	&	57.2	&	56.8	&	46.7	&	38.2	&	50.2	&	54.8	&	46.9	&	46.9	&	15.9	&	38.2	&	48.8	&	38.8	&	42.0	\\
		\hline						
	\end{tabular}
\end{table*}
\normalsize

\begin{table*}[t]
	\renewcommand{\arraystretch}{1.3}
	\caption{Path Performance Prediction Error (MAPE in \%) of NMF for Additive Metrics } \label{t_addMF}
	\centering
	\tabcolsep=0.12cm
	\renewcommand{\arraystretch}{0.9}
	\begin{tabular}{c|c|c|c|c|c|c|c|c|c|c|c|c|c|c|c|c|c|c|c|c|c}	
		\hline
		\multicolumn{2}{c|}{ }  & \multicolumn{5}{c|}{AS3967} &  \multicolumn{5}{c|}{AS3257} &  \multicolumn{5}{c|}{AS1221} &  \multicolumn{5}{c}{AS15706}\\	
		
		\hline
		\multirow{2}{*}{$\frac{|S|}{|T|}$} & sampling & \multicolumn{3}{c|}{BPR} &  \multicolumn{2}{c|}{MHR} & \multicolumn{3}{c|}{BPR} &  \multicolumn{2}{c|}{MHR} & \multicolumn{3}{c|}{BPR} &  \multicolumn{2}{c|}{MHR}   & \multicolumn{3}{c|}{BPR} &  \multicolumn{2}{c}{MHR}\\
		\cline{3-22}
		& method & UN & Re & UD & Re & UD & UN &  Re & UD  & Re & UD &  UN & Re & UD & Re & UD & UN & Re & UD & Re & UD\\
		\hline					
%
%
\multirow{2}{*}{20\%}	&	random	&	34.4	&	37.2	&	41.3	&	59.4	&	49.7	&	33.2	&	47.4	&	40.7	&	56.2	&	45.0	&	33.9	&	35.6	&	41.5	&	46.7	&	45.4	&	22.0	&	29.5	&	29.5	&	28.9	&	31.3	\\
&	monitor	&		55.9  &	55.7  &	56.4  &	57.9  &	58.3  &	61.2  &	40.7  &	60.2  &	63.3  &	60.2     &	71.2  &	37.7  &	40.5  &	69.6  &	69.4 &	69.0  &	30.8  &	33.3  &	33.8  &	69.3   \\
\hline
\multirow{2}{*}{25\%}	&	random	&	34.3	&	35.4	&	41.8	&	58.9	&	49.6	&	32.6	&	44.1	&	38.5	&	50.1	&	43.5	&	33.2	&	34.8	&	41.1	&	41.9	&	43.0	&	20.5	&	24.6	&	28.8	&	25.6	&	29.1	\\
      &	monitor	&   54.3  &	54.0  &	55.4  &	57.2  &	56.5  &	60.8  &	40.4  &	40.4  &	62.9  &	60.5   &	69.7  &	35.7  &	39.5  &	69.1  &	69.3 &	69.5  &	29.2  &	33.2  &	30.5  &	69.2  \\
\hline
\multirow{2}{*}{30\%}	&	random	&	33.8	&	35.3	&	37.0	&	55.5	&	46.6	&	31.9	&	34.9	&	36.4	&	50.0	&	42.9	&	32.6	&	33.7	&	37.1	&	38.6	&	43.1	&	19.2	&	22.2	&	26.0	&	25.1	&	28.7	\\
      &	monitor	&	53.8  &	53.7  &	54.7  &	57.6  &	56.6   &	59.5  &	39.1  &	59.1  &	61.2  &	61.9   &	68.7  &	34.8  &	36.5  &	68.6  &	69.8    &		69.4  &	28.8  &	30.3  &	28.7  &	69.6  \\
		\hline						
	\end{tabular}
\end{table*}
\normalsize

\begin{table*}[t]
	\caption{Path Performance Prediction Error (MAPE in \%) of NeuMF for Additive Metrics } \label{t_addNeuMF}
	\centering
	\tabcolsep=0.12cm
	\renewcommand{\arraystretch}{0.9}
	\begin{tabular}{c|c|c|c|c|c|c|c|c|c|c|c|c|c|c|c|c|c|c|c|c|c}	
		\hline
		\multicolumn{2}{c|}{ }  & \multicolumn{5}{c|}{AS3967} &  \multicolumn{5}{c|}{AS3257} &  \multicolumn{5}{c|}{AS1221} &  \multicolumn{5}{c}{AS15706}\\	
		
		\hline
		\multirow{2}{*}{$\frac{|S|}{|T|}$} & sampling & \multicolumn{3}{c|}{BPR} &  \multicolumn{2}{c|}{MHR} & \multicolumn{3}{c|}{BPR} &  \multicolumn{2}{c|}{MHR} & \multicolumn{3}{c|}{BPR} &  \multicolumn{2}{c|}{MHR}   & \multicolumn{3}{c|}{BPR} &  \multicolumn{2}{c}{MHR}\\
		\cline{3-22}
		& method & UN & Re & UD & Re & UD & UN &  Re & UD  & Re & UD &  UN & Re & UD & Re & UD & UN & Re & UD & Re & UD\\
		\hline					
		\multirow{2}{*}{20\%}	&	random	&	64.3  &	74.5  &	84.1  &	98.2  &	66.3  &71.9  &	39.0  &	55.8  &	48.8  &	58.9  &	65.7  &	16.1  &	38.9  &	28.0  &	47.0  &	84.5  &	\bf{4.7}  &	26.8  &	\bf{7.2}  &	30.7\\
		&	monitor	&	52.1  &	59.6  &	57.2  &	70.7  &	58.5  & 48.8  &	49.0  &	55.6  &	61.4  &	57.6     &72.5  &	23.3  &	59.6  &	35.3  &	61.7 &	47.3  &	\bf{13.9}  &	39.9  &	\bf{10.1}  &	50.2   \\
		\hline
		
		\multirow{2}{*}{25\%}	&	random	&	45.9  &	43.8  &	45.2  &	80.8  &	58.5 & 60.7  &	37.1  &	52.3  &	49.6  &	55.6 &	62.0  &	\bf{14.3}  &	36.9  &	27.0  &	46.6  &	74.6 &	\bf{4.4}  &	26.1  &	\bf{5.3}  &	27.1 \\
		&	monitor	& 45.7  &	47.9  &	53.7  &	64.0  &	55.8   &		42.5  &	46.6  &	51.2  &	55.4  &	56.9   &		65.2  &	23.0  &	50.6  &	32.4  &	57.3 &	43.5  &	\bf{8.0}  &	26.6  &	\bf{9.5}  &	37.2 \\
		\hline
		
		\multirow{2}{*}{30\%}	&	random	&	42.0  &	43.8  &	39.4  &	65.3  &	56.3  &	38.1  &	34.5  &	44.5  &	46.0  &	53.2  &	58.5  &	\bf{13.5}  &	34.9  &	23.0  &	43.0  &	46.7  &	\bf{3.9}  &	20.9  &	\bf{5.3}  &	25.0\\
		&	monitor	&	42.0  &	46.3  &	47.4  &	55.2  &	54.6  & 40.7  &	43.6  &	49.3  &	51.6  &	53.1   &	60.2  &	17.8  &	49.2  &	30.6  &	55.9    &		35.0  &	\bf{7.9}  &	26.9  &	\bf{7.8}  &	36.1  \\
		\hline						
	\end{tabular}
\end{table*}
\normalsize

\begin{table*}[t]
\renewcommand{\arraystretch}{1.3}
\caption{Path Performance Prediction Error (MAPE in \%) of NeuTomography for Non-Additive Metrics} \label{t_cong}
\centering
\tabcolsep=0.22cm
\renewcommand{\arraystretch}{0.9}
\begin{tabular}{c|c|c|c|c|c|c|c|c|c|c|c|c|c|c|c|c|c}	
\hline
\multicolumn{2}{c|}{ }  & \multicolumn{4}{c|}{AS3967} &  \multicolumn{4}{c|}{AS3257} &  \multicolumn{4}{c|}{AS1221} &  \multicolumn{4}{c}{AS15706}\\	
\hline
\multirow{2}{*}{$\frac{|S|}{|T|}$} & sampling & \multicolumn{2}{c|}{BPR} &  \multicolumn{2}{c|}{MHR} & \multicolumn{2}{c|}{BPR} &  \multicolumn{2}{c|}{MHR} & \multicolumn{2}{c|}{BPR} &  \multicolumn{2}{c|}{MHR}   & \multicolumn{2}{c|}{BPR} &  \multicolumn{2}{c}{MHR}\\
\cline{3-18}
& method &  Re & UD & Re & UD &  Re & UD & Re & UD &  Re & UD & Re & UD &  Re & UD & Re & UD\\
\hline					
\multirow{3}{*}{20\%}	&	random	&	\bf{0.7}  &	\bf{1.0}  &	35.2  &	\bf{14.5}  &	\bf{0.2}  &	\bf{1.8}  &	25.2  &	\bf{7.9}  &	\bf{1.1}  &	\bf{0.9}  &	\bf{9.7}  &	\bf{7.0}  &	\bf{0.2}  &	\bf{0.5}  &	\bf{1.8}  &	\bf{2.0}  \\
	&	monitor	&	\bf{3.3}  &	\bf{3.0}  &	57.0  &	17.9  &	\bf{0.6}  &	\bf{1.2}  &	34.2  &	\bf{11.9}  &	\bf{0.6}  &	\bf{2.1}  &	15.3  &	\bf{13.9}  &	\bf{0.3}  &	\bf{2.1}  &	\bf{8.8}  &	\bf{5.7}  \\
	&	m+PAT	&	-	&	-	&	47.8  &	18.0  &	-	&	-	&	31.6  &	\bf{11.8}  &	-	&	-	&	\bf{14.0}  &	\bf{13.0}  &	-	&	-	&	\bf{8.0}  &	\bf{5.8}  \\
\hline
\multirow{3}{*}{25\%}	&	random	&	\bf{0.8}  &	\bf{1.1}  &	28.3  &	\bf{12.3}  &	\bf{0.5}  &	\bf{1.0}  &	18.7  &	\bf{7.7}  &	\bf{0.3}  &	\bf{0.6}  &	\bf{7.3}  &	\bf{6.0}  &	\bf{0.2}  &	\bf{0.1}  &	\bf{1.2}  &	\bf{1.8}  \\
	&	monitor	&	\bf{1.6}  &	\bf{2.6}  &	53.4  &	\bf{14.4}  &	\bf{0.6}  &	\bf{1.7}  &	31.5  &	\bf{9.8}  &	\bf{0.7}  &	\bf{2.0}  &	\bf{13.3}  &	\bf{10.3}  &	\bf{0.3}  &	\bf{1.3}  &	\bf{6.0}  &	\bf{4.2}  \\
	&	m+PAT	&	-	&	-	&	53.0  &	\bf{15.0}  &	-	&	-	&	31.0  &	\bf{9.8}  &	-	&	-	&	\bf{13.5}  &	\bf{1.6}  &	-	&	-	&	\bf{6.1}  &	\bf{4.4}  \\
\hline
\multirow{3}{*}{30\%}	&	random	&	\bf{0.4}  &	\bf{1.2}  &	28.2  &	\bf{11.0}  &	\bf{0.3}  &	\bf{0.9}  &	16.1  &	\bf{8.1}  &	\bf{0.2}  &	\bf{0.5}  &	\bf{6.1}  &	\bf{5.9}  &	\bf{0.2}  &	\bf{0.3}  &	\bf{1.8}  &	\bf{1.7}  \\
	&	monitor	&	\bf{1.5}  &	\bf{2.0}  &	50.7  &	\bf{13.6}  &	\bf{0.6}  &	\bf{1.3}  &	29.7  &	\bf{8.9}  &	\bf{0.2}  &	\bf{1.7}  &	\bf{11.6}  &	\bf{9.5}  &	\bf{0.3}  &	\bf{0.8}  &	\bf{2.2}  &	\bf{3.7}  \\
	&	m+PAT	&	-	&	-	&	50.0  &	\bf{14.1}  &	-	&	-	&	29.0  &	\bf{9.0}  &	-	&	-	&	\bf{12.0}  &	\bf{9.6}  &	-	&	-	&	\bf{2.5}  &	\bf{4.0}  \\
\hline						
\end{tabular}
\end{table*}
\normalsize

\begin{table*}[t]
	\caption{\textcolor{mycolour}{Path Performance Prediction Error (MAPE in \%) of ANMI for Non-Additive Metrics}} \label{t_XXX}
	\centering
	\tabcolsep=0.225cm
	\renewcommand{\arraystretch}{0.9}
	\begin{tabular}{>{\color{mycolour}}c|>{\color{mycolour}}c|>{\color{mycolour}}c|>{\color{mycolour}}c|>{\color{mycolour}}c|>{\color{mycolour}}c|>{\color{mycolour}}c|>{\color{mycolour}}c|>{\color{mycolour}}c|>{\color{mycolour}}c|>{\color{mycolour}}c|>{\color{mycolour}}c|>{\color{mycolour}}c|>{\color{mycolour}}c|>{\color{mycolour}}c|>{\color{mycolour}}c|>{\color{mycolour}}c}	
		\hline
		& \multicolumn{4}{c|}{AS3967} &  \multicolumn{4}{c|}{AS3257} &  \multicolumn{4}{c|}{AS1221} &  \multicolumn{4}{c}{AS15706}\\	
		
		\hline
		threshold & \multicolumn{2}{c|}{BPR} &  \multicolumn{2}{c|}{MHR} & \multicolumn{2}{c|}{BPR} &  \multicolumn{2}{c|}{MHR} & \multicolumn{2}{c|}{BPR} &  \multicolumn{2}{c|}{MHR}   & \multicolumn{2}{c|}{BPR} &  \multicolumn{2}{c}{MHR}\\
		\cline{2-17}
		ratio  & Re & UD & Re & UD &  Re & UD  & Re & UD  & Re & UD & Re & UD & Re & UD & Re & UD\\
		\hline
		30\% & 15.0	&	33.0	&	24.8	&33.5	&	30.1	&	31.3	&	39.2	&	33.4	&	\bf10.3	&	32.9	&	16.9	&	34.2	&	19.6	&	35.4	&	22.4	&	36.0	\\
		50\%	&	15.4	&	46.1	&	25.1	&	52.4	&	30.1	&	49.4	&	39.1	&	51.1	&	\bf11.5	&	50.0	&	18.6	&	54.6	&	22.3	&	48.8	&	25.8	&	52.6	\\
		70\%	&	19.7	&	62.2	&	29.9	&	65.5	&	31.2	&	64.1	&	39.9	&	65.5	&	\bf12.6	&	65.5	&	20.4	&	67.4	&	23.7	&	71.1	&	26.7	&	74.1	\\
		\hline						
	\end{tabular}
\end{table*}
\normalsize


\begin{table*}[t]
\renewcommand{\arraystretch}{1.3}
\caption{Path Performance Prediction Error (MAPE in \%) of NMF for Non-Additive Metrics} \label{t_congMF}
\centering
\tabcolsep=0.20cm
\renewcommand{\arraystretch}{0.9}
\begin{tabular}{c|c|c|c|c|c|c|c|c|c|c|c|c|c|c|c|c|c}	
\hline
\multicolumn{2}{c|}{ }  & \multicolumn{4}{c|}{AS3967} &  \multicolumn{4}{c|}{AS3257} &  \multicolumn{4}{c|}{AS1221} &  \multicolumn{4}{c}{AS15706}\\	
\hline
\multirow{2}{*}{$\frac{|S|}{|T|}$} & sampling & \multicolumn{2}{c|}{BPR} &  \multicolumn{2}{c|}{MHR} & \multicolumn{2}{c|}{BPR} &  \multicolumn{2}{c|}{MHR} & \multicolumn{2}{c|}{BPR} &  \multicolumn{2}{c|}{MHR}   & \multicolumn{2}{c|}{BPR} &  \multicolumn{2}{c}{MHR}\\
\cline{3-18}
& method &  Re & UD & Re & UD &  Re & UD & Re & UD &  Re & UD & Re & UD &  Re & UD & Re & UD\\
\hline					
\multirow{2}{*}{20\%}	&	random	&	46.2	&	19.5	&	64.7	&	24.2	&	23.3	&	23.2	&	80.7	&	20.4	&	24.3	&	27.3	&	41.1	&	22.5	&	26.1	&	26.9	&	45.9	&	22.4	\\
	&	monitor	&	30.0	&	24.7	&	57.4	&	28.8	&	22.1	&	26.4	&	71.6	&	27.8	&	22.5	&	32.5	&	35.6	&	30.3	&	24.8	&	33.4	&	29.7	&	30.5	\\
\hline
\multirow{2}{*}{25\%}	&	random	&	29.5	&	16.9	&	60.1	&	23.4	&	17.2	&	21.5	&	75.0	&	18.7	&	17.4	&	26.4	&	40.1	&	20.0	&	23.4	&	26.9	&	34.2	&	22.2	\\
	&	monitor	&	29.4	&	24.9	&	54.7	&	26.4	&	22.1	&	26.3	&	62.3	&	25.2	&	22.0	&	28.6	&	35.5	&	27.0	&	24.8	&	32.5	&	28.9	&	28.7	\\
\hline
\multirow{2}{*}{30\%}	&	random	&	28.5	&	15.2	&	51.9	&	21.1	&	17.0	&	20.5	&	68.9	&	17.2	&	16.7	&	26.3	&	28.3	&	18.7	&	18.6	&	25.7	&	26.4	&	18.7	\\
	&	monitor	&	25.0	&	22.2	&	54.2	&	23.6	&	21.6	&	24.5	&	57.1	&	23.4	&	21.7	&	26.8	&	35.2	&	24.2	&	24.5	&	32.0	&	28.6	&	26.4	\\			\hline		
\end{tabular}
\end{table*}
\normalsize

\begin{table*}[t]
\vspace{.5em}
\renewcommand{\arraystretch}{1.3}
\caption{Path Performance Prediction Error (MAPE in \%) of NeuMF for Non-Additive Metrics} \label{t_congNeuMF}
\centering
\tabcolsep=0.21cm
\renewcommand{\arraystretch}{0.9}
\begin{tabular}{c|c|c|c|c|c|c|c|c|c|c|c|c|c|c|c|c|c}	
\hline
\multicolumn{2}{c|}{ }  & \multicolumn{4}{c|}{AS3967} &  \multicolumn{4}{c|}{AS3257} &  \multicolumn{4}{c|}{AS1221} &  \multicolumn{4}{c}{AS15706}\\	
\hline
\multirow{2}{*}{$\frac{|S|}{|T|}$} & sampling & \multicolumn{2}{c|}{BPR} &  \multicolumn{2}{c|}{MHR} & \multicolumn{2}{c|}{BPR} &  \multicolumn{2}{c|}{MHR} & \multicolumn{2}{c|}{BPR} &  \multicolumn{2}{c|}{MHR}   & \multicolumn{2}{c|}{BPR} &  \multicolumn{2}{c}{MHR}\\
\cline{3-18}
& method &  Re & UD & Re & UD &  Re & UD & Re & UD &  Re & UD & Re & UD &  Re & UD & Re & UD\\
\hline					
\multirow{2}{*}{20\%}	&	random	&	16.3  &	\bf{3.5}  &	55.8  &	65.9  &	\bf{7.0}  &	\bf{5.5}  &	34.7  &	79.0  &	\bf{3.6}  &	\bf{7.5}  &	46.8  &	89.4  &	\bf{5.8}  &	\bf{8.6}  &	20.7  &	34.5 \\
	&	monitor	&	15.7  &	\bf{6.2}  &	58.1  &	71.6  &	\bf{8.6}  &	\bf{9.2}  &	37.5  &	64.8  &	\bf{3.8}  &	\bf{6.5}  &	38.1  &	58.4  &	\bf{5.3}  &	\bf{7.3}  &	19.6  &	39.0 \\
\hline
\multirow{2}{*}{25\%}	&	random	&	16.0  &	\bf{4.0}  &	53.4  &	45.8  &	\bf{6.0}  &	\bf{4.5}  &	34.1  &	30.7  &	\bf{3.3}  &	\bf{5.7}  &	30.7  &	50.9  &	\bf{3.8}  &	\bf{4.1}  &	17.3  &	20.5 \\
	&	monitor	&	\bf{14.0}  &	\bf{5.8}  &	57.1  &	57.9  &	\bf{7.2}  &	\bf{8.5}  &	33.7  &	55.5  &	\bf{3.6}  &	\bf{6.4}  &	33.0  &	51.7  &	\bf{4.9}  &	\bf{6.0}  &	16.3  &	30.1 \\
\hline
\multirow{2}{*}{30\%}	&	random	&	16.8  &	\bf{3.4}  &	52.6  &	44.2  &	\bf{4.4}  &	\bf{3.6}  &	27.1  &	26.9  &	\bf{2.7}  &	\bf{2.6}  &	28.3  &	44.0  &	\bf{3.3}  &	\bf{2.1}  &	\bf{14.1}  &	18.5 \\
	&	monitor	&	\bf{12.9}  &	\bf{4.6}  &	56.8  &	51.7  &	\bf{7.6}  &	\bf{5.6}  &	33.5  &	46.7  &	\bf{3.3}  &	\bf{6.2}  &	26.1  &	49.5  &	\bf{4.0}  &	\bf{3.6} &	\bf{14.7}  &	29.2 \\
\hline						
\end{tabular}
\vspace{.8em}
\end{table*}
\normalsize


\subsubsection{Additive Metrics}
The results for additive performance metrics are shown in Table~\ref{t_add}, where MAPEs less than $15\%$ are highlighted. For each AS, $20\%$--$30\%$ node pairs are measured to infer the path performance of the rest unmeasured $70\%$--$80\%$ node pairs. In Table~\ref{t_add}, as expected, with the increased portion of measured node pairs (from $20\%$ to $30\%$), the prediction accuracy is improved for all cases.
Moreover, under random sampling, we observe that NeuTomography is exceptionally accurate, irrespective of networks, link weight distributions, or the underlying routing strategies. When $30\%$ random node pairs are measured, the corresponding MAPE ranges from $2\%$ to $7\%$ for BPR, and $5\%$--$15\%$ for MHR (with $22\%$ MAPE for AS3967 as the only exception). Furthermore, when the underlying routing strategy coincides with the performance metric of interest (i.e., BPR), such high prediction accuracy is  achievable even when only $20\%$ node pair measurements are available. By comparison, under monitor-based sampling, the resulting MAPE is relatively large, especially in the case of $|S|/|T|=20\%$. Nevertheless, when $|S|/|T|$ is increased to $30\%$, MAPE is reduced by half for many cases. On the other hand, even without increasing the amount of training data, using algorithm PAT alone improves the prediction accuracy significantly. As shown in Table~\ref{t_add}, MAPEs are almost halved (some even less than $15\%$) after applying PAT, which therefore demonstrates the high efficiency of PAT in reducing the prediction error. 

Table~\ref{t_add} also reveals some insights into the performance of NeuTomography. First, Table~\ref{t ISP} shows that link weights in AS3967 and AS3257 have higher variance comparing to those in AS1221 and AS15706, which potentially causes difficulty in predicting link performance metrics. Nevertheless, when the underlying routing mechanism is BPR, NeuTomography is robust against link weight variances and achieves high prediction accuracies for all networks. Intuitively, this is because under BPR, \textcolor{mycolour}{the routing mechanism considers both the network topology and link weight information; in other words,} 
node pair measurements incorporate more network information, including the link weight variance. Such rich information therefore enables the high prediction accuracy for NeuTomography. Second, comparing to BPR, the average MAPE under MHR is substantially larger. This is because MHR only captures the network topology information while the link weight information is lost. Nevertheless, under random sampling and UD, MAPE is generally less than $15\%$, even if the underlying routing mechanism is MHR. Furthermore, Table~\ref{t_add} implies that the effect of link weight variance becomes prominent under MHR. Specifically, MAPE is improved (even less than $3\%$) in AS1221 and AS15706 with smaller link weight variance. This observation suggests that NeuTomography is capable of dealing with different types of routing mechanisms so long as the link weight variance is relatively small. Finally, recall that links in AS1221 and AS15706 have the same weight distribution; however, we observe that the average MAPE in AS15706 is smaller, which can be explained by the network structural properties: The average node degree is $4.7$ and $5.4$ for AS1221 and AS15706, respectively. Thus, 
AS15706 is relatively densely connected, which provides more next-hop options when constructing end-to-end paths, thus leading to smaller MAPE.

We now compare our solution to the benchmarks (in Tables~\ref{t_MMPDAIL}--\ref{t_addNeuMF}). For MMP+DAIL, we know from \cite{Ma13IMC,Ma14INFOCOM} that it does not incur any error if all assumptions are completely satisfied. However, as shown in Table~\ref{t_MMPDAIL}, it is extremely vulnerable to the topology error. Specifically, when the topology error is only $0.5\%$, the MAPE can be up to $37\%$; when the topology error is increased to $2\%$, then MAPE is deteriorated to around $50\%$. This result shows the advantages of NeuTomography, for which no additional network information is required while still achieving superior performance.
In addition, under the same experiment setting,
our solution significantly outperforms NMF and NeuMF with up to one order of magnitude reduction in MAPE, which demonstrates the superiority of NeuTomography. We note that NeuMF performs relatively well for AS15706 when link metrics are from real data. This is because unlike NMF that only leverages linear transformations, NeuMF is a neural-network-based model, which is equipped with the capability in capturing non-linear relationships. 
However, even with such improvement, NeuMF is still inferior to our proposed solution. 


\vspace{.5em}
{\bf\emph{Discussions on PAT:}} In Table~\ref{t_add}, only the results of PAT for monitor-based sampling are presented. We also test PAT under random sampling and get similar results (thus omitted for page limitations). This is because PAT is proposed mainly for addressing the overfitting problem during training. For random sampling, it is unbiased in the sense that path performance metrics in the input data closely represents the performance metric distribution in the testing data. However, monitor-based sampling is biased, which causes overfitting, and PAT is able to alleviate the effect of biased sampling. 

\subsubsection{Non-additive Metrics}
For non-additive performance metrics (Table~\ref{t_cong}), with the increased portion of measured node pairs (from $20\%$ to $30\%$), the prediction accuracy is also improved. Furthermore, under BPR, NeuTomography achieves significant performance (MAPE$<3.5\%$) for all cases. This is because, regarding the performance metric of interest (congestion level), the goal for BPR is to find the least congested path. Specifically, for a node pair, if there exists a path bypassing all highly congested links, then its performance metric is small. For the tested AS topologies, most end-to-end path performance metrics fall into the narrow region of $3$--$9$, which therefore simplifies the performance inference task. Nevertheless, our objective is not only to identify the coarse-grained congestion range, but also to determine the fine-grained congestion level. In this regard, NeuTomography achieves high performance inference accuracy for each node pair without any other network information as prior knowledge. For non-additive metrics, since using only $20\%$ node pairs already enables superior performance (MAPE$<3.5\%$) under BPR, PAT is not employed for the performance improvement.

By contrast, under MHR, MAPE experiences various error levels. When link congestion levels are uniformly distributed, the corresponding MAPE is between $1.6\%$ and $18\%$ for all networks. However, under real link congestion level distribution, MAPE is large for AS3967 and AS3257, while still small for AS1221 and AS15706 ($1\%$--$15\%$). As discussed before, this is caused by the link variance and limited information in the path measurements. In Table~\ref{t ISP}, the variance of the link congestion level is severe for AS3967 and AS3257 ($78$ and $41.3$ respectively). Moreover, under MHR, the routing path is  independent of the link congestion level. Hence, when the variance of the link congestion level is large, there is a lack of link information embedded in the path measurements. Nevertheless, when the link congestion variance is relatively small, i.e., in AS1221 and AS15706, NeuTomography exhibits high accuracy. 
Furthermore, 
the large node degree in AS1221 and AS15706 also shortens the average path length under MHR, which reduces the likelihood of constructing a path with a bottleneck link, thus improving the prediction accuracy.

Besides, we also observe that monitor-based sampling generally incurs larger MAPE under MHR, especially when the link congestion level is from the real data and the networks are AS3967 and AS3257. 
Again, this observation shows that when the variance of the link congestion level is small, NeuTomography is able to recover the network information that is critical to the performance prediction even when the underlying routing mechanism is independent of the performance metric of interest. Furthermore, under MHR, we test PAT for monitor-based sampling. The results in Table~\ref{t_cong} show that when the performance metric is non-additive, PAT only slightly reduces MAPE or achieves similar performance as the one without PAT. This implies that under MHR and non-additive performance metrics, it is difficult to know the performance bound of an unmeasured node pair especially when the link congestion level exhibits high variance.

Regarding benchmark solutions, \textcolor{mycolour}{evaluation results of ANMI is shown in Table~\ref{t_XXX}.  The \emph{threshold ratio} here is the ratio of normalized $\tau$ (i.e., $\tau$ minus the minimum link metric value) and the range of the metric value distribution (i.e., the difference between the maximum and the minimum link metric values). From the results  in Table~\ref{t_XXX}, we can see that although ANMI performs relatively well for AS1221 under BPR with real link metrics, NeuTomography still offers an order of magnitude performance improvements for the same setting.} Moreover, NeuTomography outperforms NMF by up to two orders of magnitude (Table~\ref{t_congMF}). While for NeuMF, although its MAPE is less than $14\%$ under BPR (see Table~\ref{t_congNeuMF}), NeuTomography still shows one order of magnitude of improvement on average.

In sum, both results on additive/non-additive performance metrics confirm the high efficiency and applicability of NeuTomography in real networks without relying on the knowledge of additional network information (e.g., network topology) or rigorous assumptions (e.g., controllable routing), thus providing a lightweight and robust solution.

\begin{table*}[t]
\renewcommand{\arraystretch}{1.3}
\caption{Topology Reconstruction Error (FPR and FNR in \%) w.r.t. Extended Adjacency Matrix $\mathbf{A}^{(m)}$} \label{t_topo1}
\centering
\tabcolsep=0.12cm
\renewcommand{\arraystretch}{0.9}
\begin{tabular}{c|c|c|c|c|c|c|c|c|c|c|c|c|c|c|c|c|c|c|c|c|c}	
\hline
\multicolumn{2}{c|}{ }  & \multicolumn{10}{c|}{AS3967 (FPR/FNR: False Positive/Negative Rate in \%)} &  \multicolumn{10}{c}{AS3257 (FPR/FNR: False Positive/Negative Rate in \%)} \\	
\hline
\multirow{2}{*}{$\frac{|S|}{|T|}$} & sampling & \multicolumn{2}{c|}{$m=1$}  & \multicolumn{2}{c|}{$m=2$} & \multicolumn{2}{c|}{$m=3$} & \multicolumn{2}{c|}{$m=4$} & \multicolumn{2}{c|}{$m=5$}  & \multicolumn{2}{c|}{$m=1$}  & \multicolumn{2}{c|}{$m=2$} & \multicolumn{2}{c|}{$m=3$} & \multicolumn{2}{c|}{$m=4$} & \multicolumn{2}{c}{$m=5$}\\
\cline{3-22}
& method & FPR & FNR & FPR & FNR & FPR & FNR & FPR & FNR & FPR & FNR & FPR & FNR & FPR & FNR & FPR & FNR & FPR & FNR & FPR & FNR\\ 
\hline					
\multirow{3}{*}{20\%}	&	random	&	0.4	&	52.9	&	2.7	&	24.3	&	3.9	&	18.7	&	\bf3.5	&	\bf14.2	&	\bf2.4	&	\bf10.0	&	0.2	&	59.4	&	1.5	&	26.2	&	2.8	&	20.2	&	\bf2.6	&	\bf12.5	&	\bf1.7	&	\bf6.8	\\
	&	monitor	&	2.1	&	73.0	&	4.6	&	68.1	&	7.4	&	62.7	&	9.3	&	57.2	&	14.2	&	53.7	&	0.5	&	77.1	&	1.8	&	69.0	&	4.3	&	63.1	&	6.6	&	57.4	&	10.2	&	52.5	\\
	&	m+PAT	&	0.0	&	72.0	&	0.4	&	65.9	&	3.0	&	54.8	&	8.6	&	43.0	&	12.4	&	33.7	&	0.0	&	76.5	&	0.1	&	67.0	&	1.9	&	62.1	&	6.6	&	55.2	&	10.2	&	49.9	\\
\hline
\multirow{3}{*}{25\%}	&	random	&	0.5	&	38.9	&	2.1	&	21.7	&	3.1	&	16.1	&	\bf3.0	&	\bf11.3	&	\bf2.5	&	\bf9.3	&	0.2	&	50.9	&	1.3	&	20.5	&	\bf2.1	&	\bf14.6	&	\bf1.8	&	\bf9.2	&	\bf1.1	&	\bf4.9	\\
	&	monitor	&	1.6	&	67.5	&	4.2	&	58.7	&	7.8	&	52.1	&	10.9	&	50.9	&	12.3	&	50.8	&	0.2	&	71.4	&	1.5	&	60.1	&	4.0	&	53.7	&	6.6	&	49.2	&	9.5	&	42.2	\\
	&	m+PAT	&	0.0	&	72.9	&	0.5	&	59.8	&	3.1	&	46.0	&	7.7	&	32.5	&	9.9	&	23.6	&	0.0	&	71.5	&	0.2	&	59.1	&	2.3	&	49.8	&	5.9	&	40.0	&	8.2	&	32.8	\\
\hline
\multirow{3}{*}{30\%}	&	random	&	0.3	&	43.7	&	1.5	&	16.4	&	\bf1.8	&	\bf8.6	&	\bf1.2	&	\bf4.8	&	\bf0.8	&	\bf3.2	&	0.2	&	43.0	&	1.1	&	16.8	&	\bf1.7	&	\bf11.6	&	\bf1.3	&	\bf6.7	&	\bf0.8	&	\bf3.9	\\
	&	monitor	&	0.6	&	61.5	&	2.7	&	56.1	&	6.0	&	48.5	&	8.9	&	43.8	&	10.5	&	39.1	&	0.1	&	52.6	&	1.0	&	27.2	&	2.5	&	18.2	&	\bf2.7	&	\bf12.2	&	\bf2.0	&	\bf7.8	\\
	&	m+PAT	&	0.0	&	60.2	&	0.6	&	53.3	&	2.8	&	42.7	&	7.4	&	32.4	&	9.4	&	21.9	&	0.0	&	52.9	&	0.2	&	27.1	&	2.2	&	17.1	&	\bf4.0	&	\bf11.2	&	\bf2.0	&	\bf7.9	\\
\hline						
%
\hline
\hline
\multicolumn{2}{c|}{ }  & \multicolumn{10}{c|}{AS1221 (FPR/FNR: False Positive/Negative Rate in \%)} &  \multicolumn{10}{c}{AS15706 (FPR/FNR: False Positive/Negative Rate in \%)} \\	
\hline
\multirow{2}{*}{$\frac{|S|}{|T|}$} & sampling & \multicolumn{2}{c|}{$m=1$}  & \multicolumn{2}{c|}{$m=2$} & \multicolumn{2}{c|}{$m=3$} & \multicolumn{2}{c|}{$m=4$} & \multicolumn{2}{c|}{$m=5$}  & \multicolumn{2}{c|}{$m=1$}  & \multicolumn{2}{c|}{$m=2$} & \multicolumn{2}{c|}{$m=3$} & \multicolumn{2}{c|}{$m=4$} & \multicolumn{2}{c}{$m=5$}\\
\cline{3-22}
& method & FPR & FNR & FPR & FNR & FPR & FNR & FPR & FNR & FPR & FNR & FPR & FNR & FPR & FNR & FPR & FNR & FPR & FNR & FPR & FNR\\ 
\hline					
\multirow{3}{*}{20\%}	&	random	&	0.2	&	22.4	&	\bf0.5	&	\bf8.1	&	\bf1.0	&	\bf3.8	&	\bf0.8	&	\bf3.0	&	\bf0.4	&	\bf2.1	&	\bf0.1	&	\bf7.4	&	\bf0.3	&	\bf1.1	&	\bf0.8	&	\bf0.7	&	\bf0.5	&	\bf1.4	&	\bf0.2	&	\bf5.0	\\
	&	monitor	&	0.9	&	76.0	&	3.2	&	64.9	&	6.4	&	64.6	&	10.6	&	62.3	&	11.6	&	57.4	&	5.0	&	60.3	&	7.6	&	50.9	&	21.3	&	39.0	&	18.3	&	39.6	&	4.3	&	40.7	\\
	&	m+PAT	&	0.0	&	75.5	&	0.2	&	61.8	&	2.5	&	62.6	&	10.9	&	56.1	&	8.9	&	42.1	&	1.7	&	55.6	&	6.3	&	43.5	&	13.4	&	30.0	&	15.7	&	33.4	&	4.0	&	33.1	\\
\hline
\multirow{3}{*}{25\%}	&	random	&	0.1	&	21.2	&	\bf0.4	&	\bf5.6	&	\bf0.4	&	\bf3.0	&	\bf0.2	&	\bf1.1	&	\bf0.2	&	\bf0.7	&	\bf0.1	&	\bf6.6	&	\bf0.2	&	\bf0.5	&	\bf0.4	&	\bf0.6	&	\bf0.1	&	\bf0.8	&	\bf0.1	&	\bf1.8	\\
	&	monitor	&	0.4	&	67.4	&	2.3	&	54.9	&	5.7	&	54.7	&	8.5	&	55.6	&	11.6	&	49.2	&	2.5	&	47.2	&	5.4	&	24.9	&	13.2	&	20.6	&	8.9	&	20.2	&	1.6	&	18.2	\\
	&	m+PAT	&	0.0	&	65.0	&	0.3	&	49.6	&	2.0	&	43.2	&	7.7	&	32.3	&	8.6	&	26.0	&	2.4	&	46.5	&	5.0	&	21.1	&	12.0	&	18.5	&	8.3	&	15.5	&	1.4	&	16.5	\\
\hline
\multirow{3}{*}{30\%}	&	random	&	0.1	&	19.5	&	\bf0.5	&	\bf3.6	&	\bf0.6	&	\bf1.9	&	\bf0.4	&	\bf1.7	&	\bf0.2	&	\bf1.4	&	\bf0.0	&	\bf5.2	&	\bf0.1	&	\bf0.3	&	\bf0.2	&	\bf0.2	&	\bf0.2	&	\bf0.3	&	\bf0.0	&	\bf1.3	\\
	&	monitor	&	0.1	&	58.6	&	1.7	&	43.4	&	5.1	&	38.2	&	5.3	&	44.1	&	7.8	&	37.9	&	0.3	&	44.4	&	4.4	&	18.7	&	8.2	&	19.4	&	8.2	&	15.7	&	1.8	&	17.8	\\
	&	m+PAT	&	0.0	&	55.2	&	0.3	&	41.7	&	2.5	&	38.6	&	5.5	&	29.6	&	7.9	&	23.5	&	0.3	&	41.2	&	\bf4.0	&	\bf13.2	&	\bf6.1	&	\bf14.7	&	\bf7.8	&	\bf12.3	&	\bf1.6	&	\bf14.6	\\
\hline						
\end{tabular}
\end{table*}
\normalsize

\subsection{Topology Reconstruction Accuracy}
When the performance metric of interest is the minimum number of hops, we use NeuTomography to reconstruct the network topology in terms of the extended adjacency matrix. To test the reconstruction accuracy, \textcolor{mycolour}{intuitively, we can use the matrix difference $({\sum_i\sum_j|A^{(m)}_{i,j}-A'^{(m)}_{i,j}|})/{n^2}$ as the evaluation metric, where $\mathbf A^{(m)}$ and $\mathbf A'^{(m)}$ are the real and constructed extended adjacency matrices, respectively. However, since the number of links in a network is generally much smaller than ${n}^{2}$, even a full zero matrix $\mathbf A'^{(m)}$ leads to a small matrix difference. Therefore,} we use the False Positive Rate (FPR) and False Negative Rate (FNR) as the evaluation metric. Specifically, let $\mathbf A^{(m)}$ and $\mathbf A'^{(m)}$ be the real and constructed extended adjacency matrices, and $\tau$ the number of non-zero elements in $\mathbf A^{(m)}$. Then FPR is the number of non-zero elements in $\mathbf A'^{(m)}$ that are zeros in $\mathbf A^{(m)}$ over $n^2-\tau$; similarly, FNR equals the number of zero elements in $\mathbf A'^{(m)}$ that are non-zeros in $\mathbf A^{(m)}$ over $\tau$. The reconstructed network topology is accurate if both FPR and FNR are small. The corresponding results are reported in Table~\ref{t_topo1}, where both FPR and FNR less than $15\%$ are highlighted. First, for extended adjacency matrices, FPR is small for all cases as usually $\tau\ll n^2$, and thus the denominator $n^2-\tau$ is  much larger than the numerator. Second, as expected, the increased number of measured node pairs is beneficial in improving the topology reconstruction accuracy. Third, under random sampling, $\mathbf A^{(1)}$, i.e., $m=1$, is mostly inaccurate (except for AS15706). Nevertheless, when $m$ is increased to $2$, FNR is reduced by over a half. Specifically, w.r.t. $\mathbf A^{(2)}$, for both AS1221 and AS15706, FPR and FNR are less than $4\%$ when $30\%$ random node pairs are measured. Fourth, monitor-based sampling yields high topology reconstruction error; 
nevertheless, for some networks, i.e., AS15706, FNR is reduced to be less than $15\%$ via PAT. Finally, since AS1221 and AS15706 have the same link congestion level distribution, Table~\ref{t_topo1} demonstrates 
that for the denser network AS15706, even $\mathbf A^{(1)}$ is accurate. Moreover, under monitor-based sampling, PAT yields low reconstruction error for $m\geq 2$. This result implies that the topology reconstruction is more accurate in networks with high density. This is because in dense networks, there exist more node pairs which are close to each other; therefore, the probability of these close node pairs that are selected for measurements are increased, which assists the learning process in NeuTomography. For benchmarks NMF and NeuMF, they can also be used to construct $\mathbf A^{(m)}$; however, their performance is substantially worse than NeuTomography, thus omitted due to page limitations. In sum, NeuTomography provides a state-of-the-art solution to reconstruct network topologies with various granularities using only a small percentage of node pair measurements without additional network knowledge.

\section{Conclusion}\label{sec:Conclusion}
We revisited the problem of network tomography from the practical perspective. Without relying on any assumptions on network topologies, protocol support, or measurement metric properties as in the literature, we established a generic tomography framework, \emph{NeuTomography}, to infer unknown network characteristics using only end-to-end path performance metrics of selected node pairs. Next, regarding the potential overfitting problem, we proposed one algorithm that utilizes active performance bound estimation as the augmented data for iteratively improving the performance prediction accuracy. Furthermore, we investigated the feasibility of employing NeuTomography to reconstruct the network topology under the given limited measurement data. Extensive experiments using real network data show that NeuTomography is robust against network parameter errors and exhibits high prediction accuracies for both additive and non-additive performance metrics, which is up to orders of magnitude improvement over benchmark solutions. Besides, with small errors in terms of extended adjacency matrices, the reconstructed network topologies also provide vital insights to network operational optimizations.

\clearpage
\bibliographystyle{IEEEtran}
\bibliography{mybibSimplified}

\end{document}